\documentclass[12pt,prd,nofootinbib,longbibliography]{revtex4-1}
\pdfoutput=1
\usepackage{amssymb}
\usepackage{amsmath}
\usepackage{hyperref}
\usepackage{graphicx}
\usepackage{color}
\usepackage[usenames,dvipsnames]{xcolor}
\newcommand{\be}{\begin{equation}}
\newcommand{\ee}{\end{equation}}
\newcommand{\bea}{\begin{eqnarray}}
\newcommand{\eea}{\end{eqnarray}}
\newcommand{\half}{\f{1}{2}}
\newcommand{\f}[2]{\frac{#1}{#2}}

\newcommand{\pfrac}[2]{\left(\frac{#1}{#2}\right)}
\newcommand{\p}[1]{\left( #1 \right)}

\newcommand{\ket}[1]{|#1\rangle}
\newcommand{\bra}[1]{\langle #1|}
\newcommand{\braket}[2]{\langle #1|#2\rangle}

\newcommand{\Poincare}{Poincar\'e\ }

\newcommand{\sgn}{{\rm{sgn}}}

\renewcommand{\Im}{{\rm{\,Im\!}}}
\renewcommand{\Re}{{\rm{\,Re\!}}}
\newcommand{\fracL}[2]{#1/(#2)}
\newcommand{\wide}[1]{\begin{widetext}#1\end{widetext}}

\begin{document}

\title{On the Theory of Continuous-Spin Particles:\\ Helicity Correspondence in Radiation and Forces}

\author{Philip Schuster}
\email{pschuster@perimeterinstitute.ca}
\affiliation{Perimeter Institute for Theoretical Physics,
Ontario, Canada, N2L 2Y5 }

\author{Natalia Toro}
\email{ntoro@perimeterinstitute.ca}
\affiliation{Perimeter Institute for Theoretical Physics,
Ontario, Canada, N2L 2Y5 }
\date{\today}

\begin{abstract}
We have recently shown that continuous-spin particles (CSPs) have covariant single-emission amplitudes with the requisite properties to mediate long-range forces.
CSPs, the most general massless particle type consistent with Lorentz symmetry, are characterized by a scale $\rho$.
Here, we demonstrate a helicity correspondence at CSP energies larger than $\rho$, in which these amplitudes are well approximated by the familiar ones for particles of helicity 0, $\pm 1$, or $\pm 2$. These properties follow from Lorentz invariance. 
We also construct tree-level multi-emission and CSP-exchange amplitudes that are unitary, appropriately analytic, and consistent with helicity-0 correspondence.  We propose sewing rules from which these amplitudes and others can be obtained.  We also exhibit a candidate CSP-graviton matrix element, which shows that the Weinberg-Witten theorem does not apply to CSPs. 
These results raise the surprising possibility that the known long-range forces might be mediated by CSPs with very small $\rho$ rather than by 
helicity 1 and 2 particles.  
\end{abstract}

\maketitle

\newpage
\tableofcontents
\newpage

\section{Introduction}
The known long-range forces in Nature are consistently modelled by exchange of helicity~1 and 2 particles.  The absence of forces mediated by higher or lower helicities can be simply explained: Lorentz-invariance requires higher-helicity interactions to fall too fast in the infrared to mediate long-range forces 
\cite{Weinberg:1964ew}, while scalars' masses are unstable to radiative corrections.  But another type of massless particle is allowed by Lorentz symmetry --- the so-called ``continuous-spin'' particles (CSPs) \cite{Wigner:1939cj}. New CSP emission amplitudes~\cite{Schuster:2013pxj} satisfy constraints analogous to Weinberg's ``soft theorems''\cite{Weinberg:1964ew,Weinberg:1965rz,Weinberg:1964ev}, suggesting that CSPs can also consistently mediate long-range forces.  

\begin{figure}[!htbp]
\includegraphics[width=\columnwidth]{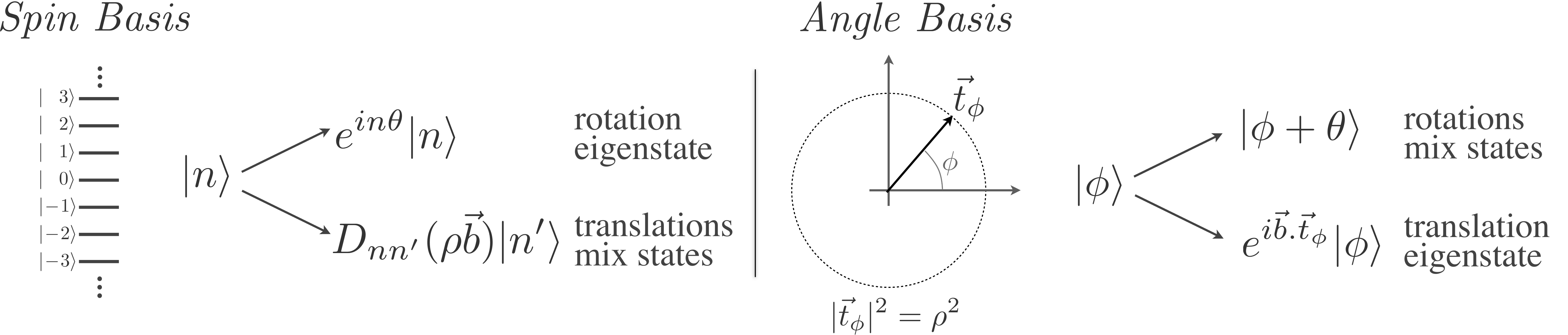}
\caption{The figure summarizes the Little Group (LG) transformation of massless particle states (see \S\ref{ssec:little}).  
Particle types are characterized by a scale $\rho$. Basis states may be labeled by a tower of integer or half-integer spins, 
or equivalently by angles on a circle.  The two bases are related by Fourier transform. 
The LG has the structure of the isometries of the Euclidean 2-plane, or $ISO(2)$. 
The spin basis diagonalizes LG rotations, while the angle basis diagonalizes LG translations.
Lorentz boosts induce LG translations (and rotations), which mix states in the spin basis. 
The scale $\rho$ controls the amount of mixing under boosts, much like the combination 
$m\times S$ for a spin-$S$ massive particle.
When $\rho=0$, spin labels become Lorentz-invariant helicities. \label{fig:spinAngle}}
\end{figure}

A CSP in 3+1 dimensions is labeled by a spin-scale $\rho$ with units of mass.  Its single-particle basis states take on arbitrary integer spins (a second type of CSP, not considered here, takes on all half-integer spins).   
Like massive particle polarizations, these spin states mix under Lorentz boosts by an amount proportional to $\rho$.
In the $\rho=0$ limit the CSP factorizes into a tower of states labeled by Lorentz-invariant integer helicities (see Figure \ref{fig:spinAngle}). 

In this paper we show, firstly, that the soft CSP emission amplitudes found in \cite{Schuster:2013pxj} smoothly approach helicity-0, 1, and 2 amplitudes in the high-energy  \emph{helicity-correspondence limit} (energy $E\gg \rho v$ for emitters at velocity $v$). 
Secondly, we provide evidence that helicity-correspondence amplitudes are consistent with tree-level unitarity and can couple to gravity, focusing on the scalar-like case.  
These findings suggest a twist on Nature's apparent preference for low helicities: in theories with CSPs, polarization modes of spin 2 or less dominate interactions at energies larger than $\rho v$.  A world with macroscopic $\rho^{-1}$ would appear, to a good approximation, to be governed at short distances by fixed-helicity gauge theories and general relativity.
If a full CSP theory of this form exists, it is conceivable that known long-range forces could be mediated by CSPs with very small (perhaps Hubble-scale) $\rho$ rather than helicity 1 and 2 particles. 
In a forthcoming paper \cite{SchusterToro:ph}, we elaborate on the thermodynamics of such a world, and estimate constraints from stellar production and radio transmission.  Further related studies will appear in \cite{Evans}.
More sharply testing this speculation requires a full classical theory of interacting CSPs, a topic to be pursued in \cite{Schuster:2013pta}.  

In Section \ref{sec:CSP_reps} we review the kinematics of CSPs and the main results of \cite{Schuster:2013pxj}: new wavefunctions and ``soft factors'' \eqref{eq:CSPsoft} into which single CSP-emission amplitudes should factorize when the CSP becomes soft compared to other interacting particles. 

Section \ref{sec:correspondence} exhibits the \emph{helicity correspondence} of these soft factors, which arises in the high-energy limit $E\gg \rho$ or, for non-relativistic emitters with velocity $v$, $E\gg \rho v$ (a more precise definition of the correspondence limit including angular dependence is $\rho |z| \ll 1$, with $z$ given by \eqref{zdefinition}).  
 We define helicity-$h$ correspondence to mean that spin-$h$ emission amplitudes differ from standard helicity-$h$ amplitudes by $(\rho|z|)^2$-suppressed corrections, while amplitudes for other spins are suppressed by increasing powers of $\rho |z|$.
We present scalar-, photon-, and graviton-like soft factors in the family \eqref{Soft_spinBasis} that exhibit helicity 0, $\pm 1$, and $\pm 2$ correspondence respectively.  We show that charge conservation and the equivalence principle, required to obtain a helicity correspondence, follow from  perturbative unitarity of the CSP interactions.  
This hierarchical structure permits approximate thermal equilibrium for the correspondence state and any matter it couples to without thermalizing the full CSP tower \cite{SchusterToro:ph}, settling Wigner's concern about CSP thermodynamics \cite{Wigner:1963}.
We exclude higher-helicity correspondence in theories with a cutoff scale much larger than $\rho$, while raising the possibility that spins may interact democratically in theories with an ultraviolet cutoff near $\rho$.  A related conjecture for double-valued CSPs was made in \cite{Iverson:1971vy}.  

In Section \ref{sec:infinite-spin}, we explore the connection between continuous-spin particles and the simultaneous high-spin limit of a massive particle, with $(mass)\times(spin)=\rho$ held fixed.  The group-theory of CSPs is closely related to this high-spin limit \cite{Inonu:1953sp}; the CSP soft factors allow us to also compare their dynamics.  The relation of soft factors to this infinite-spin limit provides some intuition for how the hierarchical coupling structure required by correspondence can be consistent with Lorentz-invariance.  But the connection is purely formal, not physical --- finite-spin matrix elements do not approach CSP matrix elements in the infinite-spin limit. This result underscores that the connection we highlight between helicity-$h$ particles and CSPs in the small-$\rho$ limit is much closer and more physical than that between CSPs and the massive high-spin limit. 

Section \ref{sec:unitarity} examines the consistency of scalar-like CSP soft factors with unitarity.  
We construct unitary tree-level multi-emission and exchange amplitudes mediated by an off-shell CSP that factorize into scalar-like soft factors, and propose candidate sewing rules for a full perturbative $S$-matrix.  Because the amplitudes and sewing rules are not rational functions of momenta, they are less constrained by factorization than familiar tree-level amplitudes.  Studying the loop-level unitarity of these sewing rules is an important problem that may sharpen either the structure of a CSP theory or potential physical obstructions.  
The interplay of unitarity and covariance for photon-like or graviton-like CSPs is more subtle than in scalar-like theories, just as it is for photons and gravitons compared to scalars.  Analogous constructions in these cases are an important direction for future work; we highlight several subtleties particular to the construction of photon- and graviton-like CSP amplitudes.

No simple analogue of the high-helicity Weinberg-Witten theorem \cite{Weinberg:1980kq} forbids CSP couplings to gravity, as discussed in \ref{sec:CSPgravity}.  We exhibit a covariant, symmetric, and conserved rank-two tensor matrix element between single-CSP states, in contrast to the non-existence of such matrix elements for high-helicity particles.  These matrix elements also have a correspondence limit in which they approach familiar stress-energy tensors for a scalar or gauge boson, though this correspondence is governed by the graviton momentum and possesses some unusual characteristics.  
It is still not clear whether these matrix elements arise from a conserved tensor \emph{operator} or whether they are compatible with CSP-matter interactions.  
Of course, a helicity-2-like CSP might yield consistent gravitational theories even if such obstructions are found, and merits further study.

\section{Review of Continuous-Spin Particles}\label{sec:CSP_reps}

This section summarizes properties of continuous-spin particles.
The classification of particle degrees of freedom that transform  as
unitary irreducible representations (irreps) of the \Poincare group has proved
a valuable principle for understanding theories of nature. Continuous-spin particles are naturally defined from this perspective.  
We review the definition, state-space, and kinematics for continuous-spin particles (\SS\ref{ssec:little}), covariant ``wavefunctions''  on which Lorentz 
transformations induce transformations appropriate to a single-particle state (\SS\ref{ssec:wavefuncs}), and the form of soft emission factors required by 
Lorentz-invariance (\SS\ref{ssec:softFactors}).  All of the material in this section is expanded upon in \cite{Schuster:2013pxj}.  

\subsection{\Poincare Transformations on One-Particle States and the Little Group}\label{ssec:little}
Because the translation generators $P^\mu$ mutually commute, one-particle states can be labeled by  a
$c$-number momentum eigenvalue $k^\mu$ and by internal labels  $a$ whose detailed form we will constrain shortly.  
We may write the action of a Lorentz transformation $\Lambda$ on each state as
\be
U(\Lambda) \ket{k,a} =\sum_{a'}  D(\Lambda,k)_{aa'} \ket{\Lambda k, a'} \label{LorentzTransf},
\ee
where the transformation matrix $D$ must be unitary with respect to the norm
\be
\braket{k,a }{k', a' } = 2 k^0 \delta^{(3)}(k-k') \delta_{aa'} \label{norm}. 
\ee
For \eqref{LorentzTransf} to be satisfied, the $a$ labels must furnish an irreducible representation of the Little Group $LG_k$, the subgroup of Lorentz transformations 
such that $\Lambda^\mu_\nu k^\nu = k^\mu$.  We exhibit some properties of this group (focusing on the massless case), then we will return to \eqref{LorentzTransf} 
in the case of general Lorentz transformations. 

The Little Group $LG_k$ is generated by the components of 
\be
w^\mu \equiv  \half \epsilon^{\mu\nu\rho\sigma} k_{\nu} J_{\rho\sigma}, \label{wkdef}
\ee
closely related to the Pauli-Lubanski pseudo-vector.  Because $w.k=0$, $w^\mu$ has three independent components and the Little Group in 3+1 dimensions is always 3-dimensional.  
These components obey the commutation relation 
\be
\left[w^\mu,w^\nu\right] = - i \hbar \epsilon^{\mu\nu\rho\sigma} w_\rho k_\sigma. \label{wwcommutator}
\ee
The operator $w^2 = w^\mu w_\mu$, with dimensions of $mass^2$,
is a Casimir of the \Poincare group, and therefore constant on all irreps of the Little Group.  
In the case of a massive particles, we see that for $k^\mu = (m, 0, 0, 0)$, $w^\mu = (0, m {\vec{\bf J}})$; the components have group structure $SO(3)$ and in the spin-$S$ representation
\be
w^2 = - m^2 \vec{\bf J}^2 = -m^2 S(S+1).
\ee
Indeed, it is reassuring that the left- and right-hand sides of this equation are manifestly covariant, unlike the three-vector expression in the middle. 

For null $k^\mu$, we introduce a light-cone frame with $k$ as one of its axes (see \cite{Heinonen:2012km} for a similar treatment).  We choose space-like $\epsilon_{1,2}(k)$ with $\epsilon_i(k)^2 = -1$, 
$\epsilon_1(k).\epsilon_2(k) = \epsilon_i(k).k = 0$, and the unique vector $q^\mu(k)$ satisfying $q(k)^2=0, q(k).k=1$, and $q(k).\epsilon_i(k)=0$.  
It will also be useful to work with $\epsilon_\pm^\mu(k) \equiv (\epsilon_1^\mu(k) \pm i \epsilon_2^\mu(k))/\sqrt{2}$.  In terms of this frame, we can 
identify the components of $w^\mu$ and their commutation relations as
\bea
R \equiv q.w \quad T_{1,2} \equiv \epsilon_{1,2}.w, & \qquad & (T_{\pm} \equiv \sqrt{2} \epsilon_\pm.w), \label{LGrotTrans}\\
\ [R,T_{1,2}] = \pm i T_{2,1}, \quad [T_1,T_2] = 0, & \qquad & ([R,T_\pm] = \pm T_\pm, \quad [T_\pm,T_\mp]=0).
\eea
For example, taking 
\be
\bar k^\mu=(\omega,0,0,\omega), \quad \epsilon_1 = (0,1,0,0), \epsilon_2 = (0,0,1,0), \label{kref}
\ee
we find
\be
w^\mu = - \bar k^\mu R + \hat \epsilon_1^\mu T_1 + \hat \epsilon_2^\mu T_2
\ee
with $R = J_{12}$, $T_1 = \omega (J_{32}+J_{02})$, and $T_2 = -  \omega (J_{31}+J_{01})$.

These are the commutation relations of the rotation and translation generators of $ISO(2)$.  The Casimir in this case is
\be
w^2 = W^2 =  - \vec{T}^2 =  - T_+ T_-, \label{w2massless}
\ee
which is completely independent of the action of the rotation generator!  
Any Little Group element can be labeled by  an angle $\theta$ and a two-vector $\vec b$ (or a complex number $\beta = (b_1+i b_2)/\sqrt{2}$) with units 
of length through the decomposition
\be
W(\theta,\beta)\equiv e^{i\,\vec b.\vec T} e^{-i\theta R} =  e^{\frac{i}{\sqrt{2}}(\beta T_{-} + \beta^* T_+)} e^{-i\theta R}.\label{LGelem}
\ee

This group permits two types of single-valued unitary representation (we will not discuss double-valued representations, though these also exist).  The helicity representations 
consist of a single state $\ket{k,h}$ transforming as
\be
W(\theta,\vec b) \ket{k,h} = e^{i h \theta} \ket{k, h}.
\ee
Since these states are annihilated by the translation generators, they have vanishing $w^2$.  The so-called ``continuous-spin'' representations of $ISO(2)$ are characterized 
by $w^2 = -\rho^2$ for any $\rho$ with dimensions of momentum.  These have infinitely many states, which can be parametrized in two particularly useful bases: simultaneous 
eigenstates of $T_{1,2}$ (the angle basis, with states labeled by an angle $\phi$) or eigenstates of $R$ (spin basis, with states labeled by an arbitrary integer $n$).  
These are simply related by Fourier transform.  We begin with the former basis which has much simpler Little Group transformation rules, then transform to the latter, 
which recovers a tower of all integer-helicity particles in the $\rho\rightarrow 0$ limit. 

Simultaneous eigenstates of $T_{1,2}$, subject to $W^2 = -\rho^2$, have the interpretation of plane-waves in the $ISO(2)$ of fixed ``momentum'' norm $\vec T^2 = \rho^2$.  
They can be labeled by an angle $\phi$ such that
\be 
\vec T \ket{k,\phi} = (\rho \cos\phi, \rho \sin\phi) \ket{k,\phi}, \quad R \ket{k,\phi} = i \partial_\phi \ket{k,\phi}.
\ee
Generic Little Group elements transform the states as 
\bea
W(\theta, \vec b) \ket{k,\phi} &= &e^{i \vec b.\vec
  t_{\phi+\theta}} \ket{k, \phi+\theta}\\
  &=& e^{i \rho \Re[\sqrt{2}\beta e^{-i(\phi+\theta)}]} \ket{k, \phi+\theta} \label{eq:LittleGroupAction}\\
  &=& \int \frac{d\phi'}{2\pi} D_{\phi\phi'}[\theta,\beta]\ket{k,\phi'}, \quad \mbox{ with } D_{\phi\phi'}[\theta,\beta] = (2\pi) \delta(\phi'-\phi-\theta) e^{i \rho \Re[\sqrt{2}\beta e^{-i\phi'}]},
\eea
which is unitary with respect to the inner product
\be
\braket{k,\phi}{k',\phi'}=2\pi 2k^0\delta^{3}(k-k')\delta(\phi-\phi').
\ee
The Lorentz-invariant sum over states is 
\be
\int \frac{d^3\vec{k}}{k^0}\frac{d\phi}{2\pi}.
\ee

We may transform to the spin basis by defining 
\be
\ket{k,n} \equiv \int \frac{d\phi}{2\pi} e^{in\phi} \ket{k,\phi},
\ee
for integer $n$, which have inner product
\be
\braket{k',n'}{k,n}= \delta_{n n'} 2k^0\delta^{3}(k-k').
\ee
Little Group transformations act as
\bea
W[\theta, \beta] \ket{k,n} & = & D_{nn'}[\theta,\beta] \ket{k,n'} ,\\
D_{n n'}[\theta,\beta] & = & e^{-in\theta}(ie^{i\alpha})^{(n-n')}J_{n-n'}(\rho\sqrt{2}|\beta|), \label{nbasisMixing}
\eea
where $\beta=|\beta|e^{i\alpha}$.
The appearance of Bessel functions is to be expected, as they are representation functions for the Euclidean 
group in two dimensions (see for example \cite{Tung:1985na} or \cite{VilenkinKlimyk}). 
We call this a ``spin'' rather than ``helicity'' basis because they mix under general Little Group actions, in contrast with helicity states.  In the limit 
$\rho\rightarrow 0$, however, $J_{n-n'}(\rho|2\beta|)$ approaches zero for $n\neq n'$ and $1$ for $n=n'$, so we recover a direct sum of all 
integer-helicity states with the transformation rule
\be
D_{n n'}[\theta,\beta] \rightarrow e^{-in\theta}\delta_{n n'}. \label{rho0limit}
\ee

So far we have considered only the action of little-group elements that keep a momentum $k^\mu$ invariant.  The action of other Lorentz 
transformations on single-particle states is dictated partly by convention.  It is standard to construct all states in a given \Poincare irrep from the 
states at \emph{fixed} reference momentum $\bar k^\mu$, as follows.  For each $k$ we choose a ``standard Lorentz transformation'' $B_k$ such that
$(B_k)^\mu_\nu \bar k^\nu =k^{\mu}$, for which we \emph{define} 
\be
D(B_k,k)_{a a'} \equiv \delta_{a a'}. \label{stdAction}
\ee
The action of any Lorentz transformation $\Lambda$ on one-particle states is then  determined by group composition of \eqref{eq:LittleGroupAction} and \eqref{stdAction} to be  
\be
U(\Lambda)\ket{k,a} = D(W_{\Lambda,k},k)_{a a'} \ket{\Lambda k, a'} 
\quad \mbox{with} \quad W_{\Lambda,k} \equiv  B_{\Lambda k}^{-1} \Lambda B_k \in LG_{\bar k},
\label{particleDef}.  
\ee
We will find it useful to think of states at given momentum $k$ as having their ``own'' Little Group.  This is equivalent to the $B_k$ construction, 
\emph{provided} that we choose, for each $k^\mu$, a light-cone frame with
$\epsilon_\pm^\mu(k) \equiv {B_k}^\mu_\nu \epsilon_\pm^\nu(\bar k).$  Since the action of $B_k$ on $\bar k^\mu$ and $\epsilon_\pm^\mu(\bar k)$ fully specifies 
a generator $B_k$, we can view any choice of $\epsilon_\pm(k)$ as implicily defining the ``standard Lorentz transformation'' $B_k$.  

It will be desirable to work in a basis where, for every $k$, $\epsilon_\pm^0(k)=0$.  Among other things, in this basis $R$ is the usual ``helicity operator'', 
${\bf \hat k}.{\vec {\bf J}}$ (the non-covariance of this operator could have been our first hint that no symmetry guarantees Lorentz-invariance of massless particles' helicities).  
One choice of $B_k$ that guarantees this, starting from \eqref{kref},
is
\be
B_k = e^{i \phi J_z}   e^{i \theta J_y}  e^{i \log(|{\bf k}|/\omega) K_z} \mbox{ for } 
k = |{\bf k}| (1, \sin\theta\cos\phi, \sin\theta \sin\phi, \cos\theta).
\label{goodStandardBoost}
\ee

Generalizations of continuous-spin representations in higher dimensions, with supersymmetry, and in other contexts are discussed in \cite{Brink:2002zx,Khan:2004nj,Mourad:2005rt,Bekaert:2005in,Edgren:2005gq,Zoller:1991hs}

\subsection{Wavefunctions}\label{ssec:wavefuncs}
Although Wigner long ago formulated covariant wave equations for continuous-spin particles \cite{Wigner:1939cj,Bargmann:1946me,Wigner:1947,Bargmann:1948ck,Wigner:1963}, they are not the most general wave equations whose solutions transform as CSPs.  As Wigner himself noted several times, wave equations and \Poincare representations are not associated uniquely, and 
we can expect multiple alternative wave equations to describe the same type of particle excitation.  
Nonetheless, most of the literature on CSPs for the last seven decades \cite{Yngvason:1970fy,Chakrabarti:1971rz,Abbott:1976bb,Hirata:1977ss} relies almost exclusively on Wigner's equations of motion (one exception is \cite{Iverson:1971hq}). 

In \cite{Schuster:2013pxj}, we sought families of wavefunctions possessing both Lorentz and Little Group labels, on which the actions of the two were related by the covariance equation  
\be
\sum_{a'} D_{a a'}\left[W_{\Lambda,k}\right] \psi\left(\mathbf{\Lambda k},a', l \right) 
= \sum_{\bar l} D_{l \bar l}^{-1}[\Lambda] \psi\left(\mathbf{k},a, \bar l \right),
\label{wfnCovariance}
\ee
where $l$, $\bar l$ and $a$, $a'$ are Lorentz and little-group indices respectively.  In the case of infinite-dimensional single-valued representations of the Lorentz group, 
one is led to seek out solutions whose Lorentz transformation is carried not by a tensor index, but by an auxiliary vector $\eta^\mu$, with
\be
D[\Lambda] \psi(\eta,\dots) \equiv \psi(\Lambda \eta, \dots),
\ee
or similarly for an auxiliary Weyl spinor field $\xi^\alpha$ \cite{Gelfand}.  While the latter approach was pursued by \cite{Iverson:1971hq}, we have found the auxiliary vector form to be 
and at least equally useful, even though the resulting Lorentz representations are generically not irreducible.  In this form, and using the angle basis for the little group, 
the covariance equation \eqref{wfnCovariance} becomes 
\be
\int \frac{d\phi'}{2\pi} \psi \left( \{ \mathbf{\Lambda k},\phi' \},\eta^\mu \right) D_{\phi' \phi}\left[W(\Lambda,k)\right]
= \psi \left( \{ \mathbf{k},\phi \}, \Lambda^{-1} \eta \right). \label{covariancePhi}
\ee
It is sufficient to solve this equation for $\Lambda \in LG_{k}$, which is readily done by linearizing the Lorentz and Little Group actions of the three generators:
\bea
-i\left( \eta.\epsilon_{-} \epsilon_{+}.\partial_{\eta}-\eta.\epsilon_{+} \epsilon_{-}.\partial_{\eta} \right) \psi &=& \partial_{\phi} \psi \\ \label{Reom}
-\left( \eta.\epsilon_{-}k.\partial_{\eta} - \eta.k\epsilon_{-}.\partial_{\eta} \right) \psi &=& \frac{\rho}{\sqrt{2}} e^{-i\phi} \psi \\ \label{T-eom}
\left( \eta.\epsilon_{+}k.\partial_{\eta} - \eta.k\epsilon_{+}.\partial_{\eta} \right) \psi  &=& \frac{\rho}{\sqrt{2}} e^{i\phi} \psi \label{T+eom}
\eea
These equations are homogeneous in $\eta$, are Fourier-conjugate to themselves, and imply (using $k^2=0$) 
\be
W^2 \psi = \left(2k.\eta \, k.\partial_{\eta}\, \eta.\partial_{\eta} - (k.\eta)^2\partial_{\eta}^2 - \eta^2(k.\partial_{\eta})^2 \right)\psi = -\rho^2 \psi.
\ee

The covariance equations above have two classes of solutions, depending on whether they have singular or smooth support near $\eta.p = 0$.  The first, singular class can be written as 
\be
\psi(\{k,\phi,f\},\eta)=\int dr f(r) \int d\tau \delta^{4}(\eta-r\epsilon(k\phi)-r \tau k)e^{-i\tau\rho}, \label{planewave}
\ee
with $k^2=0$, $f(r)$ arbitrary, and 
\be
\epsilon(k\phi) \equiv \frac{i}{\sqrt{2}}( \epsilon_{+}(k)e^{-i\phi}-\epsilon_{-}(k)e^{i\phi})= -\sqrt{2} \Im \left[ \epsilon_+ e^{-i\phi}\right].\label{eq:epsilonkphi}
\ee
These solutions are supported on a plane in $\eta$-space, with a profile under $\eta$-rescaling determined by $f(r)$.  Because they satisfy $p^2\psi = p.\eta \psi=0$, 
the $W^2$ equation simplifies to $\left( -\eta^2(p\cdot\partial_{\eta})^2+\rho^2 \right) \psi = 0$.  These, together with the equation $(\eta^2+1)\psi=0$, are the Wigner 
equations, which are satisfied by $f(r)=\delta(r-1)$ (in which case $\psi$ is only supported on a line in $\eta$).  However, alternate  $\eta$-space equations such as 
$\eta.\partial_\eta \psi=0$ single out different $f(r)$ but yield equally covariant wavefunctions.  

Smooth solutions of the covariance equations \eqref{Reom}--\eqref{T+eom} take the form 
\bea
\psi(\{k,\phi\},\eta) &=& f(\eta.k,\eta^2) e^{i\rho \frac{\eta.\epsilon(k\phi)}{\eta.k}} \nonumber \\
&=&   f(\eta.k,\eta^2) e^{-i\rho\sqrt{2} \Im \left( \frac{\eta.\epsilon_{+}(k)e^{-i\phi}}{\eta.k} \right) } \label{smoothWavefunction}
\eea
in the angle basis, with $f$ an arbitrary function.  
Two further wave equations besides $k^2 \psi =0$ and $(W^2 +\rho^2)\psi = 0$ must be imposed to fully fix $f$.  One simple choice is $k.\partial_\eta \psi = 0$, 
$\eta.\partial_\eta \psi = m\psi$, which fixes $f(\eta.k,\eta^2) = (\eta.k)^m$.  
Equivalently, the smooth wavefunctions in the spin-basis are 
\be
\psi(\{k,n\},\eta)  =  f(\eta.k,\eta^2) e^{i n \arg\left[\frac{- \eta.\epsilon_{+}(k)}{\eta.k}\right] } J_{n}\bigg(\rho \sqrt{2}\bigg|\frac{\eta.\epsilon_{+}(k)}{\eta.k}\bigg|\bigg). \label{wfn_spinBasis}
\ee

The subclass of wavefunctions supported on $f(\eta.k,\eta^2)= \delta(\eta^2) (\eta.k)^c$ for arbitrary complex $c$ are closely related to the wavefunctions found by \cite{Iverson:1971hq}. For a discussion of the relation between the above wavefunctions and the equations of motion for high-helicity particles (appropriate when $\rho=0$) \cite{Fronsdal:1978rb,Sorokin:2004ie}, see \cite{Schuster:2013pxj}.

\subsection{Soft Factors for CSP Emission}\label{ssec:softFactors}
Together, Lorentz-invariance, locality, and unitarity impose significant constraints on how different massless particles can interact. These constraints are particularly simple 
in the case of amplitudes involving $n$ particles of momentum $p_1,\dots, p_n$, plus a single massless ``soft particle'' whose momentum $k$ satisfies  
$k.p_i \ll p_i.p_j$ for all $i,j$.  Famously, Weinberg found these limits to be so constraining as to exclude any coupling of helicity $h>2$ particles that would mediate a 
long-range force, and to imply charge-conservation (the equivalence principle) in helicity-1 (2) interactions \cite{Weinberg:1964ew}.

In the limit $k.p_i\rightarrow 0$, unitarity implies that these amplitudes are dominated by ``external emission'' terms where the soft particle is emitted by one of the in- or 
out-going hard particles.  The sum of these contributions takes the form 
\be
A(p_1,\dots,p_n, \{k,a\}) = A_0(p_1,\dots,p_n) \times \left[ \sum_{i=1}^n \frac{1}{\pm 2 p_i.k + i\epsilon} \times s_i(\{k, a\},p_i)\right] + {\cal O}(|k|^0), \label{softFactorization}
\ee
where $a$ denotes the Little Group state of the soft particle, and the upper (lower) sign in the propagator corresponds to outgoing (incoming) momenta.
In this expression, the ``parent amplitude'' $A_0$, which involves the $n$ hard particles but not the soft particle, is universal, while each external emission off the $i$'th leg is proportional to a distinct ``soft factor'' $s_i$.  
Simultaneous Lorentz-covariance of the parent $n$-point amplitude and the $n+1$-point soft amplitude imply a very simple covariance relation on the object in square brackets:
\bea
f(\{k, a\},p_1,\dots,p_n) & = & \sum_{a'} D_{a a'}^{*}[W_{\Lambda, k}] f(\{\Lambda k, a'\},p'_1,\dots,p'_n) + {\cal O}(|k|^0),\label{softCovariance}\\
\mbox{where } f(\{k, a\},p_1,\dots,p_n) & \equiv & \sum_{i=1}^n \frac{1}{\pm 2 p_i.k + i\epsilon} \times s_i(\{k, a\},p_i).
\eea
This condition is most simply satisfied if the soft factors $s_i$ are separately covariant, though cancellations between terms are possible (and indeed required in the case of 
helicity amplitudes, which is the origin of the constraints in \cite{Weinberg:1964ew}).

In \cite{Schuster:2013pxj}, we exhibited a simple soft-factor ansatz that satisfies \eqref{softCovariance} term by term, obtained simply by evaluating the smooth covariant 
wavefunctions \eqref{smoothWavefunction} at $\eta^i = p^i$: for CSP \emph{emission} this gives  
\bea
s_i(\{k,\phi\},p_i) & = & f_i(k.p_i) e^{-i \rho \frac{\epsilon(k\phi).p_i}{k.p_i}} = f_i(k.p_i) e^{+i \sqrt{2} \rho \Im\left[e^{-i\phi} \frac{{\epsilon^+}.p_i}{k.p_i}\right]}, \mbox{ and } \label{eq:CSPsoft} \\
s_i(\{k,n\},p_i)  & = & f_i(k.p_i) \tilde J_n\bigg(\rho \sqrt{2}\frac{\epsilon_{+}.p_i}{k.p_i}\bigg)\label{Soft_spinBasis}
\eea
in the angle and spin bases, respectively, with 
\be
\tilde J_{n}(w) \equiv (-1)^n e^{-i n \arg(w)} J_n(|w|).\label{Jtilde}
\ee
CSP absorption soft factors are obtained by complex conjugation of the above.  To highlight the helicity correspondence, we will use the spin basis in most of this paper.  But we remind the reader that most non-trivial manipulations of the soft factors are much simpler if one first Fourier transforms them back to the angle basis.  

We have allowed for the possibility that each species $i$ has a different $f_i$, and omitted the explicit dependence on $\eta^2 \rightarrow p_i^2 = m_i^2$, since this is independent of kinematics for on-shell $p$.  Whatever their form, the $f_i$ must have mass dimension 1.  Since $f$ must be smooth as $k.p_i \rightarrow 0$, a natural decomposition of $f_i$ is into terms
\be
 f^{(m)}(k.p) = g^{(m)}(k.p)^m.
\ee
where $g^{(m)}$ has mass-dimension $1-2m$.  

\section{Helicity Correspondence in Soft Emission}\label{sec:correspondence}
This section and the next explore \emph{helicity correspondence} in a variety of CSP scattering amplitudes.  We have already seen from \eqref{rho0limit} a kinematic connection between CSPs and fixed-helicity particles: in the $\rho \rightarrow 0 $ limit, the (single-valued) continuous-spin representation decomposes into a tower of states with all integer helicities.  Helicity correspondence is a stronger condition on interacting CSP theories, namely that the CSPs' \emph{interactions} have a well-defined (and non-trivial) $\rho\rightarrow 0$ limit, in which they approach fixed-helicity interactions.  The deviation from helicity-like amplitudes at finite $\rho$ is controlled by $\rho/(energy)$ on dimensional grounds.  Thus for any finite $\rho$, the physical \emph{high-energy} limit is also helicity-like.  The precise scaling of the correspondence parameter $z_i \equiv - \sqrt{2} \epsilon^*_+(k).p_i/k.p_i$ in different kinematic regimes is discussed in \ref{ss:scalarCSP}.

In this section, we explore helicity correspondence in the CSP soft factors \eqref{Soft_spinBasis}.  Because soft factors are almost fully constrained by Lorentz-invariance, we will be able to classify their correspondence systematically.  At the same time, because generic amplitudes in a unitary theory of CSPs (if one exists) must factorize into soft factors, we expect this correspondence to persist in more general amplitudes of any full theory. As we will see in the next section, the construction of more general CSP amplitudes is not always straightforward.  Nonetheless, soft factor correspondence will be a useful guide to exploring physical consequences of correspondence in a universe with CSPs \cite{SchusterToro:ph}. 

The pattern of correspondence exhibited by CSP soft factors, summarized in Table \ref{tab:soft}, has three striking features:
\begin{itemize}
\item Although the $\rho \rightarrow 0$ \emph{spectrum} contains all integer helicities, only helicities 0, $\pm 1$, and $\pm 2$ can interact in the correspondence limit.  There exist choices of $f(p_i.k)$ in \eqref{Soft_spinBasis} for which \emph{only} helicity 0, helicity $\pm 1$, or helicity $\pm 2$ interact with matter in the $\rho\rightarrow 0$ limit.  We refer to these as helicity-$h$ correspondence soft factors, or as scalar-, photon-, or graviton-like soft factors respectively.  
\item The soft factors recovered in the $\rho\rightarrow 0$ limit reproduce the leading interactions of scalars, gauge bosons, and gravitons coupled to charges (it could have been that, for example, we only recovered dipole interactions of gauge bosons proportional to $F_{\mu\nu}$, rather than the soft factors associated with $A_\mu J^\mu$ couplings in gauge theories).  This is especially surprising because unlike the CSP soft factors \eqref{Soft_spinBasis}, helicity-1 and 2 soft factors are not Lorentz-covariant.  
\item Though we define correspondence through a $\rho\rightarrow 0$ limit, the soft factors exhibiting correspondence  --- including the requirements of charge conservation for helicity 1 and 2 interactions --- are singled out by considerations internal to the physics of a CSP with fixed $\rho$!  The criterion that selects helicity-0 and 1 correspondence is the absence of an ultraviolet cutoff imposed by the tree-level interactions.  Helicity-2-like interactions must of course have a cutoff (the Planck scale) but these are the interactions with the \emph{weakest} cutoff-dependence, after the helicity-0 and 1 cases.
\end{itemize}

\begin{table}[htbp]
\begin{tabular}{@{}l@{\hspace{0.2in}}l@{\hspace{0.2in} }l@{\hspace{0.2in} }l}
\toprule
 & Scalar-like & Photon-like & Graviton-like \\
\hline
\vspace{0.05in}
Helicity Soft Factor & 
$1$ & 
$q_i \epsilon_\pm(k).p_i $ & 
$\frac{1}{M_*} (\epsilon_\pm(k).p_i)^2 $\\
\vspace{0.05in}
\parbox[t]{1.3in}{\raggedright CSP Soft Factor (covariant form)}& 
$a_i \tilde J_n(\rho z_i)$ & 
$\frac{- q_i}{\sqrt{2} \rho} p_i.k \tilde J_n(\rho z_i)$  & 
$\frac{1}{\rho^2 M_*} [(p_i.k)^2 - \tfrac{\rho^2}{4} p_i^2] \tilde J_n(\rho z_i)$\\
\hline
\vspace{0.05in}
Constraint & 
--- & 
$\sum_{in} q_i = \sum_{out} q_i$ & 
universal $\tfrac{1}{M_*}$ \\
\vspace{0.05in}
Subtraction & 
--- & 
$\frac{- q_i}{\sqrt{2} \rho} p_i.k \delta_{n0}$ & 
$\frac{1}{\rho^2 M_*}p_i.k (p_i.k \delta_{n0}+ \rho \epsilon_+.k \delta_{n1}-\rho \epsilon_-.k \delta_{n,-1})$\\
\botrule
\end{tabular}
\caption{\label{tab:soft}Soft factors for helicity $\pm h$ and CSP soft factors that exhibit correspondence; soft factor amplitudes are written in terms of $z_i \equiv \sqrt{2} \epsilon_+^*(k).p_i/k.p_i$ and $\tilde J_n(w) \equiv (-1)^n e^{-i n \arg(w)} J_{n}(|w|)$.  The constraint in the third row for helicities $\pm 1$ and $2$ is required in the helicity case to ensure Lorentz covariance of the amplitude and in the CSP case to ensure perturbative unitarity (at least up to the scale $M_*$).  When the constraint is satisfied, the ``subtraction'' in the fourth row sums to zero over all legs of any amplitude, ensuring both perturbative unitarity and correspondence of the CSP emission amplitudes.  The combination (covariant CSP soft factor) -- (subtraction) also defines a completely equivalent (albeit non-covariant) \emph{subtracted} CSP soft factor, which in the limit of small $z$ approaches the helicity soft factor.}
\end{table}

Each of the features emphasized above will be made precise and justified in this section.  We recall in \S\ref{ss:preliminarySoft} the relationship of soft factors to scattering amplitudes and the form of gauge and gravity soft factors --- these preliminaries form a necessary baseline for the discussion of helicity correspondence.  In \S\ref{ss:scalarCSP} we demonstrate the helicity-0 correspondence of the simplest CSP soft factors, and elaborate on the kinematic parameter that controls deviations of CSP soft emission amplitudes from the helicity-like form.  Sections \ref{ss:photonCSP} and \ref{ss:gravitonCSP} introduce soft factors with helicity-1 (photon-like) and helicity-2 (graviton-like) correspondence, and show how charge conservation conditions are related to the perturbative unitarity of high-energy scattering.  This discussion resolves the puzzle that CSP soft factors are Lorentz-covariant while gauge/gravity soft factors are not, and motivates the introduction of equivalent but non-covariant CSP soft factors that tighten the connection to gauge and gravity soft-factors.  
We step back in Section \ref{ss:generalCSP} to see why the most general CSP soft factor has \emph{only} scalar-like, photon-like, and graviton-like correspondence.  

\subsection{Preliminaries: Soft Factors and Amplitudes}\label{ss:preliminarySoft}
Before exploring correspondence in CSP soft factors, we review the standard soft factors for helicities 0, 1, and 2 that we will recover from CSPs in the $\rho \rightarrow 0$ and high-energy limits.  In addition, we introduce a scalar-scattering example in which single-emission amplitudes at generic momenta are simply related to \eqref{softFactorization2}, not just approaching it in a limiting sense.  This example will allow us to exhibit correspondence in a simple high-energy limit, rather than the more subtle double limit of high energy and soft emission, to which we return in \S\ref{ss:generalCSP}.

Soft factors are defined through the limit $k.p_i \ll p_i.p_j$ of a single-emission amplitude \\
$A(p_1,\dots,p_n, \{k,a\})$.  In this limit, the ($n$+1)-point amplitude $A$ must factorize as
\be
A(p_1,\dots,p_n, \{k,a\}) = A_0(p_1,\dots,p_n) \times \left[ \sum_{i=1}^n \frac{1}{\pm 2 p_i.k + i\epsilon} \times s_i(\{k, a\},p_i)\right] + {\cal O}(|k|^0), \label{softFactorization2}
\ee
where $A_0$ is a parent amplitude of $n$ ``matter'' legs, and the $s_i$ are soft factors that depend only on the soft momentum $k$ and the $i$'th matter leg.  In the case of helicity 0, $\pm 1$, and $\pm2$, the soft factors are
\be
s_i(\{k, h=0\},p_i) = a_i, \quad s_i(\{k, h=\pm 1\},p_i) = q_i \epsilon_\pm(k).p_i, \quad s_i(\{k, h=\pm 2\},p_i) = g (\epsilon_\pm(k).p_i)^2 \label{helicitySoft}
\ee
where $a_i$, $q_i$, and $g$ have mass-dimension 1, 0, and -1 respectively.  The latter two soft factors are not individually Lorentz-covariant.  Covariance of \eqref{softFactorization2} therefore requires that the sum of incoming charges $q_i$ equals the sum of outgoing charges (helicity $\pm 1$) and that $g$ is universal (helicity $\pm 2$), so that the non-covariant terms in the transformations of \eqref{helicitySoft} sum to zero~\cite{Weinberg:1964ew}.

\begin{figure}[!htbp]
\includegraphics[width=0.85\columnwidth]{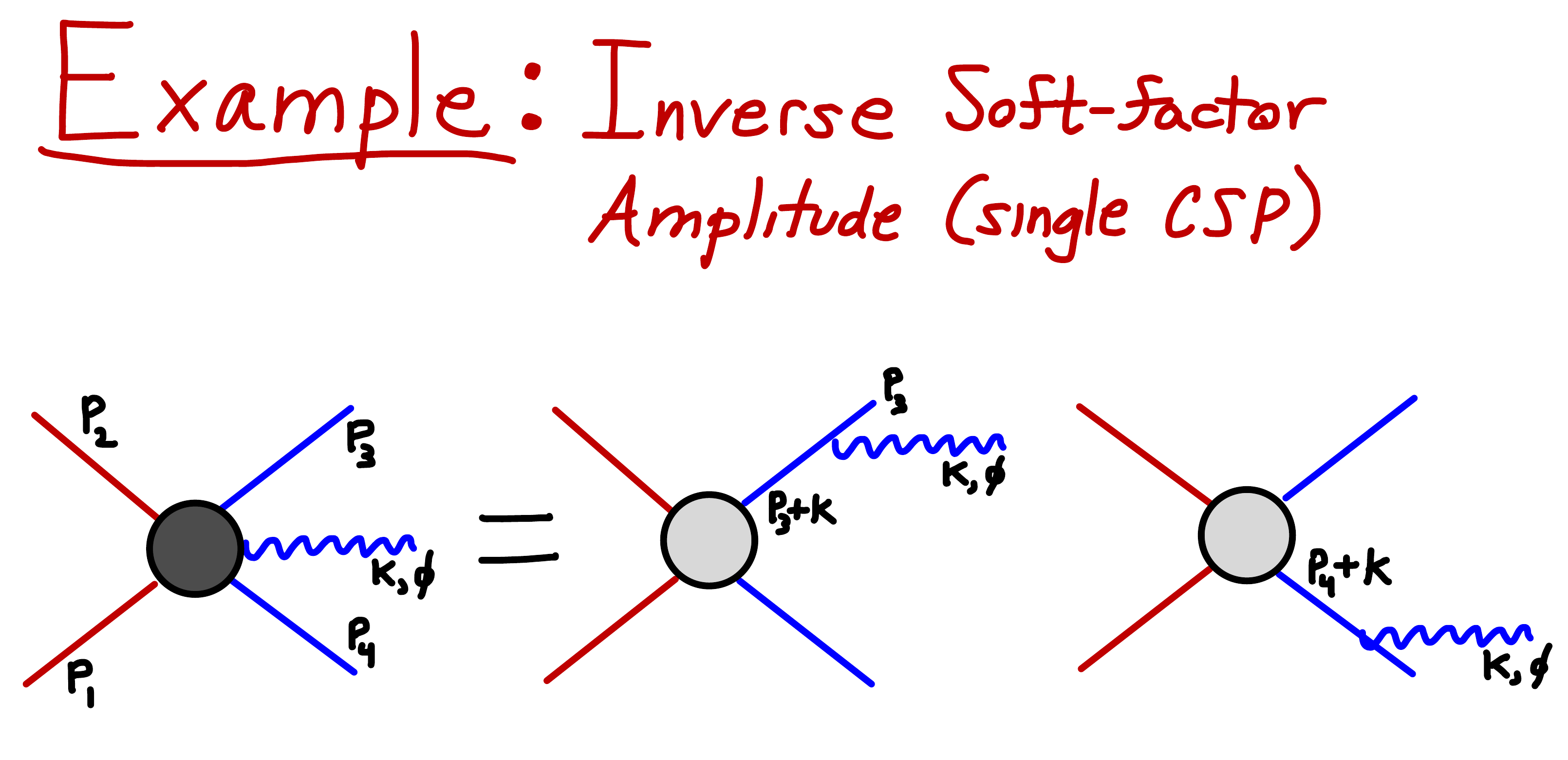}
\caption{The construction of a candidate on-shell CSP amplitude is illustrated above. 
CSPs are attached to a parent amplitude $A_4(p_1,p_2; p_3,p_4)$ using the CSP soft factors, 
with appropriate matter propagators included. This example presumes that only the outgoing 
matter legs couple to the CSP. The final result is the 5-point amplitude $A(p_1,p_2; p_3,p_4,\{k, \phi \})$
used below as an example to investigate certain aspects of CSP interactions. \label{fig:single-CSP}}
\end{figure}

For generic matter legs and interactions, \eqref{softFactorization2} is only unitary and Lorentz-invariant in the $k \rightarrow 0$ limit --- at finite $k$, poles in $A_0$ must be shifted and the sum over matter spins becomes non-trivial.  However, for scalar matter with momentum-independent parent amplitude $A_0$ (i.e.  matter legs interacting only through a non-derivative contact interaction), \eqref{softFactorization2} is a valid amplitude for all $k$.  For example,
if $p_1 \dots p_4$ correspond to distinct particle types with $A_4(p_1, p_2 \rightarrow p_3, p_4) = \lambda$, 
\bea
A(p_1,p_2 \rightarrow p_3,p_4,\{k, a \}) = \lambda  & \bigg( &
\frac{s(\{k, a \}, p_1)}{(k-p_1)^2+i\epsilon} + \frac{s(\{k, a \}, p_2) }{(k-p_2)^2+i\epsilon} \\
& & + \frac{s(\{k, a \}, p_3)}{(p_3+k)^2+i\epsilon}
+ \frac{s(\{k, a \}, p_4) }{(p_4+k)^2+i\epsilon}\bigg). \label{eq:singleParticle}
\eea
This should be interpreted as the lowest O($\lambda$) contribution, at tree level. 
For the special case that only the outgoing particles couple to the radiated particle of momentum $k$, this is illustrated diagrammatically in Figure \ref{fig:single-CSP}.  This formula implicitly defines $s(\{k, a \}, p_i)$ at arbitrary $k$ (not just in the soft limit), which we will continue to refer to as a ``soft factor'' in the following discussion.  We will return to the original definition of soft factors, as the \emph{dominant} contributions to the $k\rightarrow 0$ limit of a generic amplitude, in \S \ref{ss:generalCSP}.

\subsection{Scalar Correspondence}\label{ss:scalarCSP}
The general soft factor \eqref{Soft_spinBasis} involved an arbitrary pre-factor $f_i(k.p_i)$ of mass-dimension 1.   
The simplest example of correspondence arises for constant $f_i(k.p_i) = a_i$, yielding
\be
s_i^{(0)}(\{k,n\},p_i)  = a_i \tilde J_n(\rho z_i)\label{eq:CSP0soft}
\ee
with
\be
z_i \equiv \sqrt{2} \epsilon^*_+(k).p_i/k.p_i, \qquad \tilde J_n (w) \equiv (-1)^n e^{-i n \arg(w)} J_{n}(|w|).\label{zdefinition}
\ee
Importantly, differential emission cross-sections obtained from \eqref{eq:singleParticle} using this soft factor are finite, because
\bea
\sum_n |s_i^{(0)}(\{k,n\},p_i)|^2  & = & |a_i|^2 \\
\sum_n s_i^{(0)}(\{k,n\},p_i) s_j^{(0)}(\{k,n\},p_j)^* & =&  a_i a_j^* J_0(\rho |z_i - z_j|).\label{softInterference}
\eea
These results follow from Bessel identities, or even more simply by working in the angle basis (in this case, the sums become  integrals of pure phases over a finite interval, which are clearly bounded).  

\begin{figure}[htbp]
\includegraphics[width=0.49\textwidth]{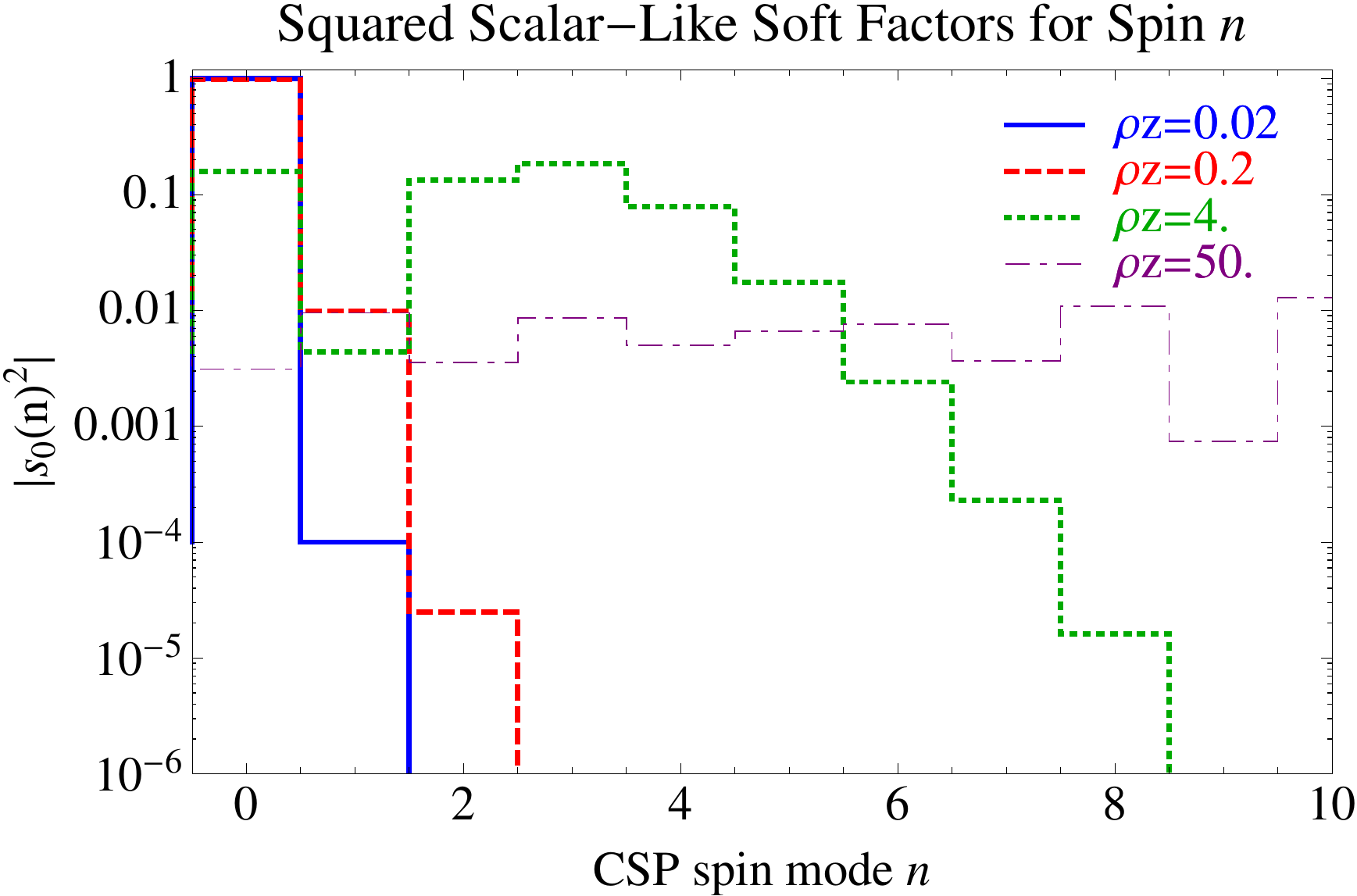}
\includegraphics[width=0.49\textwidth]{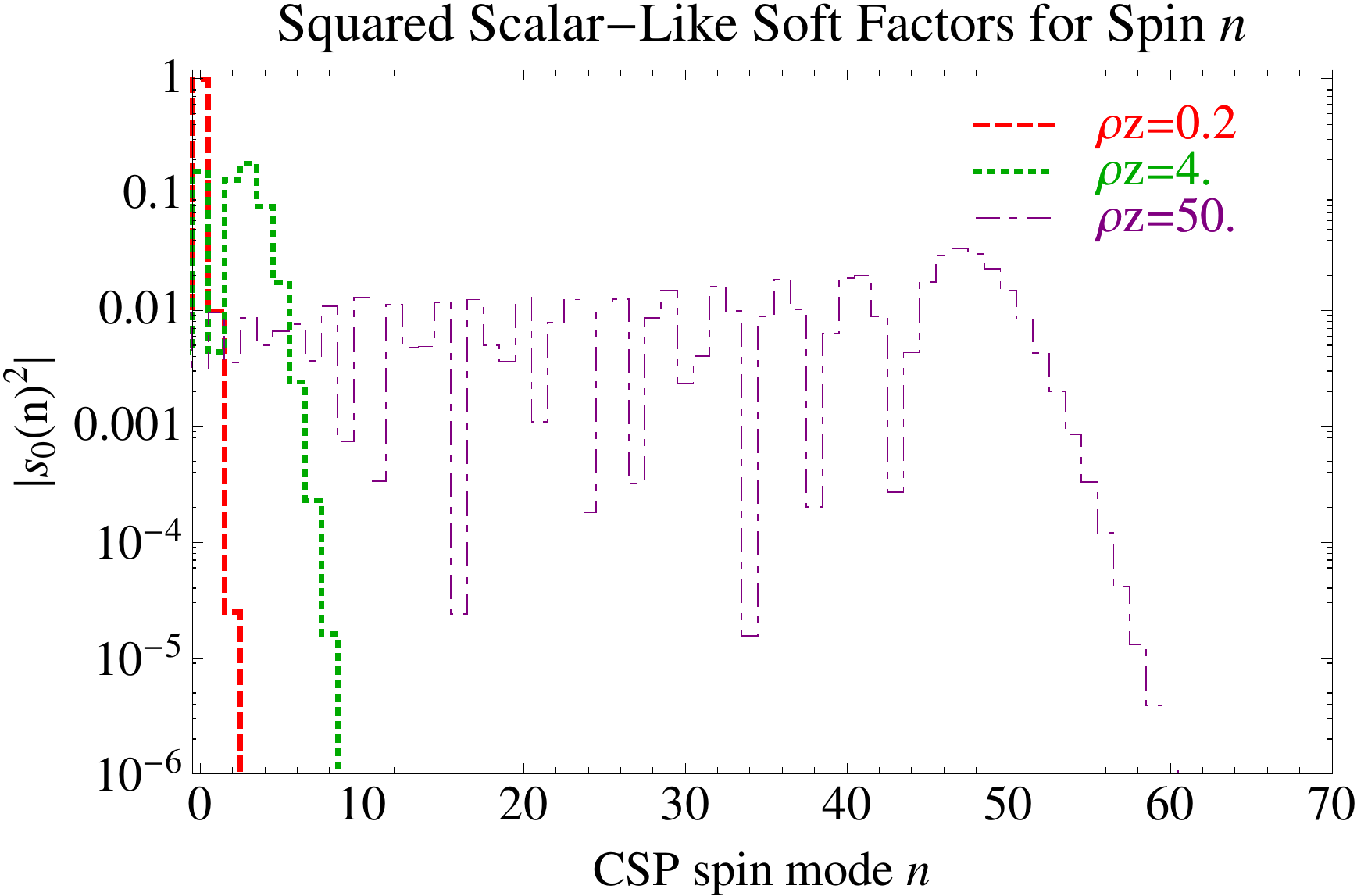}
\caption{\label{fig:scalarLikeZplots}Both plots illustrate on a log scale the scaling of squared scalar-like soft factors $|s^{(0)}(\{k,n\},p)|^2$ with $\rho |z| \equiv \sqrt{2} \rho \left|\frac{\epsilon_+(k).p}{k.p}\right|$.  The left plot focuses on low $n$ modes, while the right plot zooms out to show the large-$n$ scaling of the soft factors for large $z$.  For small $\rho |z|$, the amplitudes are sharply peaked at $n=0$.  For large $\rho |z|$, spin-$n$ soft factors for $n\lesssim \rho |z|$ are all $O(1/\sqrt{\rho z})$, while $n \gtrsim \rho |z|$ modes are further suppressed.  Negative $n$ modes scale the same way as positive $n$.}
\end{figure}

We turn now to the limit $\rho |z_i| \ll 1$, where helicity-0 correspondence is recovered.  Taylor-expanding \eqref{eq:CSP0soft} at small $z_i$, we find 
\be
s_i^{(0)}(\{k,n\},p_i) \approx  a_i \left(\frac{\rho z_i}{2}\right)^n \left( 1-\frac{\rho^2 |z_i|^2}{4 (n+1)} + \dots \right)/n!. \label{eq:CSP0softCorresp} 
\ee
for $n\ge 0$ and its complex conjugate for $n<0$.  
Thus for $\rho |z_i| \ll 1$, 
the dominant matter-CSP coupling is to the $n=0$ spin-mode of
  the CSP, and is well approximated by the helicity-0 soft factor $s_i(\{k,n\},p_i) = a_i$.  
The emission amplitudes for other modes are suppressed by
$|\rho z_i|^n$.  This is illustrated in Figure \ref{fig:scalarLikeZplots}, which shows the squared soft factors for the spin-$n$ modes for a range of small and large values of $\rho z$.

This is our first example of correspondence.  We will shortly elaborate on the kinematic dependence of the dimensionful parameter $|z_i|$ that controls the correspondence, but comment first on its frame-dependence.  Our  $z_i$ depends explicitly on the choice of frame vectors $\epsilon_{\pm}$ introduced in \S\ref{ssec:little}, or equivalently on the choice of a standard boost $B_p$.  This is expected, because the CSP spin-states are not distinguished in a Lorentz-invariant way, and mix under boosts.  It is similar to the frame-dependence encountered in polarization amplitudes for massive particles.  Here, we can gain intuition for the frame-dependence by recalling that the spin label $n$ dictates how the state re-phases under the {\it little group} rotation defined by \eqref{LGrotTrans}. For generic $\epsilon_+(k)$, the LG rotation is a linear combination of a Lorentz rotation and a Lorentz boost.    Only when $\epsilon_\pm^0=0$ in a particular frame does the LG rotation coincide with a pure Lorentz rotation about the CSP 3-momentum direction $\hat {\bf k}$, $R=\frac{\vec{\mathbf k}\cdot \vec{{\mathbf J}}}{|{\mathbf k}|}$, making contact with the standard definition of helicity and with $SO(3)$ rotation properties of amplitudes.  The case $z=0$ is realized by an emitter at rest \emph{for the choice of $B_p$ where $R$ is a pure rotation}.  
This is why the standard boost \eqref{goodStandardBoost} with $\epsilon_+^0(k)=0$ for all $k$ is particularly useful.

But there must be a Lorentz-invariant notion of correspondence limits that applies to any Lorentz-invariant quantity.  In general, these will take the form of $\rho |z_i - z_j|$, which is clearly invariant under both shifts of $\epsilon_+ \rightarrow \epsilon_+ + k$ (which cancel between the two $z$'s) and re-phasings of $\epsilon_+$, and therefore under changes of frame.  For example, though the $n=0$ emission amplitude off a particular scattering reaction is not Lorentz-invariant, the total emission probability is.  For the CSP-emission process shown in Figure \ref{fig:single-CSP} this is 
\bea
 \sum_n |A(p_1,p_2 \rightarrow p_3,p_4,\{k, n \})|^2 &=& |\lambda|^2  \left|
\frac{s^{(0)}(\{k, n\}, p_3)}{(p_3+k)^2+i\epsilon} + \frac{s^{(0)}(\{k, n\}, p_4) }{(p_4+k)^2+i\epsilon}\right|^2\\
& = & |\lambda|^2  \bigg(
\frac{|a_3|^2}{((p_3+k)^2)^2} + \frac{|a_4|^2}{((p_4+k)^2)^2} \nonumber \\
 && \qquad \qquad + \frac{2 Re[a_3 a_4^*] J_0(\rho |z_i - z_j|)}{(p_3+k)^2(p_4+k)^2}\bigg),\label{probability0}
\eea
where $a_{3,4}$ are coupling constants and we have used \eqref{softInterference} to obtain the second line.  When $\rho |z_i - z_j| \ll 1$, the Bessel function is approximately 1 and we recover the total emission probability of a scalar theory; when $\rho |z_i - z_j|$ is large, the interference terms drop out and the emission pattern (and total emission probability) changes significantly.  A related example, which we will discuss further in \cite{SchusterToro:ph}, is the emission and absorption of a CSP by two different radiators. There too, deviations from a helicity theory are controlled by $\rho |z_{emitter} - z_{absorber}|$.  It will always be simplest, however, to work in either the rest frame of one particle of interest or a center-of-mass frame for high-energy processes, so that even frame-dependent quantities such as soft factors exhibit correspondence.  

We now consider the physical meaning of small $\rho z_i$, for both relativistic and non-relativistic $p$.  Indeed, $z_i$  is a rather familiar expression that appears in the Klein-Nishina formula and other photon emission problems.  
As was discussed previously, a convenient choice of $B_k$ is
\eqref{goodStandardBoost}, for which $\epsilon_+^0(k) =0$ for all $k$.  Then if $p=(p^0,{\mathbf
  p})$ and ${\bf \vec k}$ is at a relative angle of $\theta$, 
\be
|z| \approx \frac{|{\mathbf p}|
  \sin\theta}{|{\mathbf k}| (p^0 - |{\mathbf p}| \cos\theta)}.
\ee
For non-relativistic $p$ with velocity $v_p$, this approaches $\fracL{v_p\sin\theta}{{|\mathbf k}|}$.In the relativistic, non-collinear limit $|z|\approx \fracL{\cot(\theta/2)}{{|\mathbf k}|}$.  The latter expression has a collinear singularity, which for massive $p$ is regulated by the matter particle's mass. For
fixed $|{\mathbf k}|$, $|z|$ is maximized at emission angles
$\sin\theta = m/E_p$ and bounded above by $\frac{|{\mathbf
    p}|}{|{\mathbf k}| m}$. 

Having considered the case of a constant $f_i(k.p_i) = a_i$ in  \eqref{Soft_spinBasis}, a natural next step is to consider simple powers $f_i(k.p_i) = (k.p_i)^m c_i^{(m)}$, where the $c_i$ have mass-dimension $1-2 m$.  We would not have considered these forms at all for helicity-0 particles, because they vanish in the $k\rightarrow 0$ limit.  However, two novel features of CSP soft factors --- their dependence on a scale $\rho$ and the $z$-dependent Bessel function/phase --- allow non-trivial soft limits even with $m>0$.  As we will see below, $m=1$ and 2 will, under suitable conditions, lead to helicity-1-like and helicity-2-like CSP interactions.

\subsection{Photon Correspondence}\label{ss:photonCSP}

On a first look, it seems that all of the soft factors would also be scalar-like in the correspondence limit, albeit with momentum-dependent interactions.  As an example, the soft factor with $f$ linear in $p_i.k$ is  
\be
s_i^{(1)}(\{k,n\},p_i)  = \frac{- q_i}{\mu} k.p_i \tilde J_n(\rho z_i),\label{eq:CSP1soft}
\ee
where we have written the coefficients in terms of a uniform mass scale $\mu$ and dimensionless coefficients $q_i$.  In analogy to \eqref{eq:CSP0softCorresp}, we find 
\be
s_i^{(1)}(\{k,n\},p_i) \approx  \frac{- q_i}{\mu} k.p_i \left(\frac{\rho z_i}{2}\right)^n/n! \left( 1-\frac{\rho^2 |z_i|^2}{4(n+1)} + \dots \right) 
\ee
for $n\ge 0$, and the conjugate for $n<0$.  
The leading interaction is still that of the spin-0 state, but now its coefficient grows with the momenta $p_i$ and $k$.  The $n=\pm 1$ soft factors are proportional to photon soft factors $\epsilon^*_\pm.p$ (since $z\propto \epsilon_+^*.p_i/k.p_i$ cancels the $p_i.k$ pre-factors), but they are still $\rho z$-suppressed relative to spin-0 interactions.  

If we were to multiply this soft factor into a four-scalar amplitude as in \eqref{eq:singleParticle} or Figure \ref{fig:single-CSP}, the spin-0 emission amplitude would be simply 
\be
A(p_1,p_2 \rightarrow p_3,p_4,\{k,n=0\}) \approx \frac{- \lambda }{ 2 \mu } (q_1 + q_2 - q_3 - q_4) + O(\rho^2|z_i|^2),\label{softCSP1amplitude}
\ee
where the explicit pre-factors in each soft factor have been cancelled against the propagators (dropping $i\epsilon$'s).  The absence of any propagator suppression to this amplitude implies that the resulting 2-to-3 scattering cross-section grows at large center-of-mass energy, becoming strongly coupled at a scale $\Lambda \approx (\mu/\lambda) (q_1+q_2-q_3-q_4)^{-1}$.  Explicitly, integrating \eqref{softCSP1amplitude} over the phase-space of the  $2$ original final-state particles \emph{and} the final-state phase-space of the CSP,
\be
\sigma_{soft}(s) \sim \frac{\lambda^2}{s} \int^{\rho<|k|<p_i} \frac{d^3 k}{2 k}
\left(\frac{q_1 + q_2 - q_3 - q_4}{\mu}\right)^2 \sim \left(\frac{q_1 + q_2 - q_3 - q_4}{\mu}\right)^2.\label{CSP1xsec}
\ee
This does \emph{not} fall as $1/s$, and therefore violates perturbative unitarity at center-of-mass energy $\sqrt{s} \sim\Lambda$.  

The only way to remove the cutoff in \eqref{softCSP1amplitude} and \eqref{CSP1xsec} with $s_i^{(1)} \neq 0$ is if $q_1+q_2=q_3+q_4$, or more generally if ``charge'' (defined as the coefficient $q_i$ of the soft factor for CSP emission off the $i$'th leg of an amplitude) is conserved in all interactions.  This ``charge conservation'' requirement ensures that the $O(\rho^0)$ terms in all $n=0$ single-emission amplitudes cancel.  Thus $n=0$ amplitudes begin at $O(\rho^2)$, and the $O(\rho^1)$ amplitude for $n=\pm 1$ dominates in the small-$\rho z$ correspondence limit:
\wide{
\be
 A(\{k,n=\pm1\},p_1\dots) = \lambda \sum_{i} q_i \frac{\rho}{\sqrt{2} \mu} \frac{\epsilon_{\pm}^*.p_i}{(\pm p_i+k)^2} \left( 1 -  |\rho z_i|^2/8 + \dots\right).\label{m1n1soft}
\ee}
For $|\rho z_i| \ll 1$, this approaches the soft-emission amplitude for a gauge boson coupling to each leg with strength $e\, q_i$ where $e=\fracL{\rho}{\sqrt{2}\mu}$.  
The next largest interactions are for $n=0$ and $\pm 2$,
\bea
 A(\{k,n=0\},p_1\dots) &=& +\frac{\lambda}{2} \sum_{i} e q_i \frac{\epsilon_-^*.p_i}{(\pm p_i+k)^2} \times \rho z_i 
\left(1- \rho^2 |z_i|^2/16 + \dots\right).\label{m1n0soft}\\
 A(\{k,n=2\},p_1\dots) &= & -\frac{\lambda}{4} \sum_{i} e q_i \frac{\epsilon_+^*.p_i}{(\pm p_i+k)^2} \times \rho z_i 
\left(1- \rho^2 |z_i|^2/12 + \dots\right).\label{m1n2soft}
\eea
More generally, the couplings for $|n|\neq 1$ scale as $e (\rho z/2)^{\left||n|-1\right|}/|n|!$, and hence vanish in the small-$z$ limit.

We have already noted the apparent inconsistency of a covariant CSP soft factor approaching the (non-covariant) gauge theory soft factor as $\rho \rightarrow 0$.  However, recall that had we considered a single soft factor \eqref{eq:CSP1soft} rather than a full amplitude, with $\mu /\rho$ fixed as $\rho\rightarrow 0$ to keep the effective ``helicity-1'' coupling $e$ fixed, the $n=0$ soft factor actually diverges!  Although this is a perfectly consistent soft factor for finite $\rho$, and the resulting amplitudes have a smooth $\rho\rightarrow0$ limit that exhibits helicity 1 correspondence, the soft factor itself has no finite $\rho\rightarrow 0$ limit.  

\begin{figure}[htbp]
\includegraphics[width=0.49\textwidth]{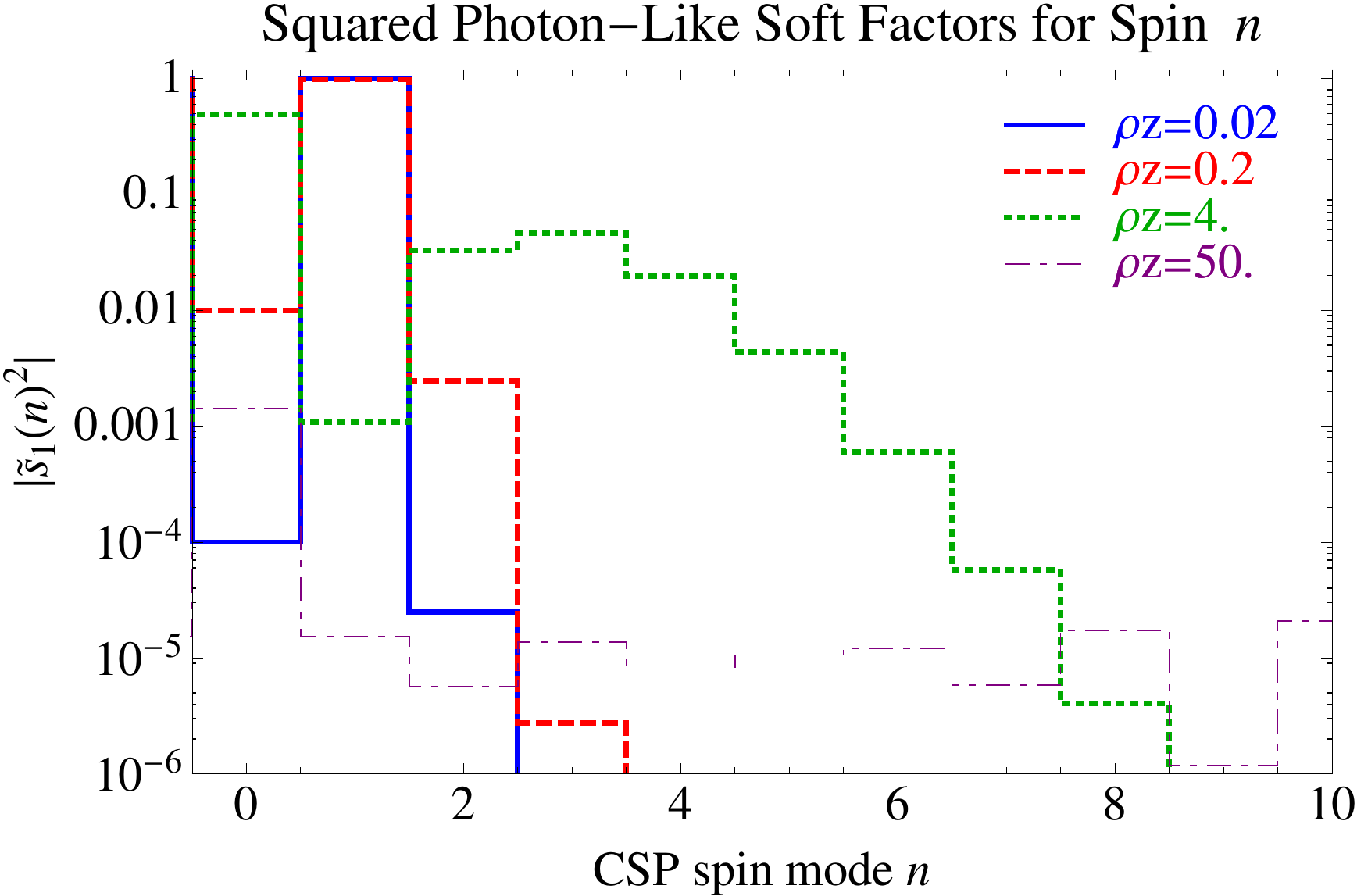}
\includegraphics[width=0.49\textwidth]{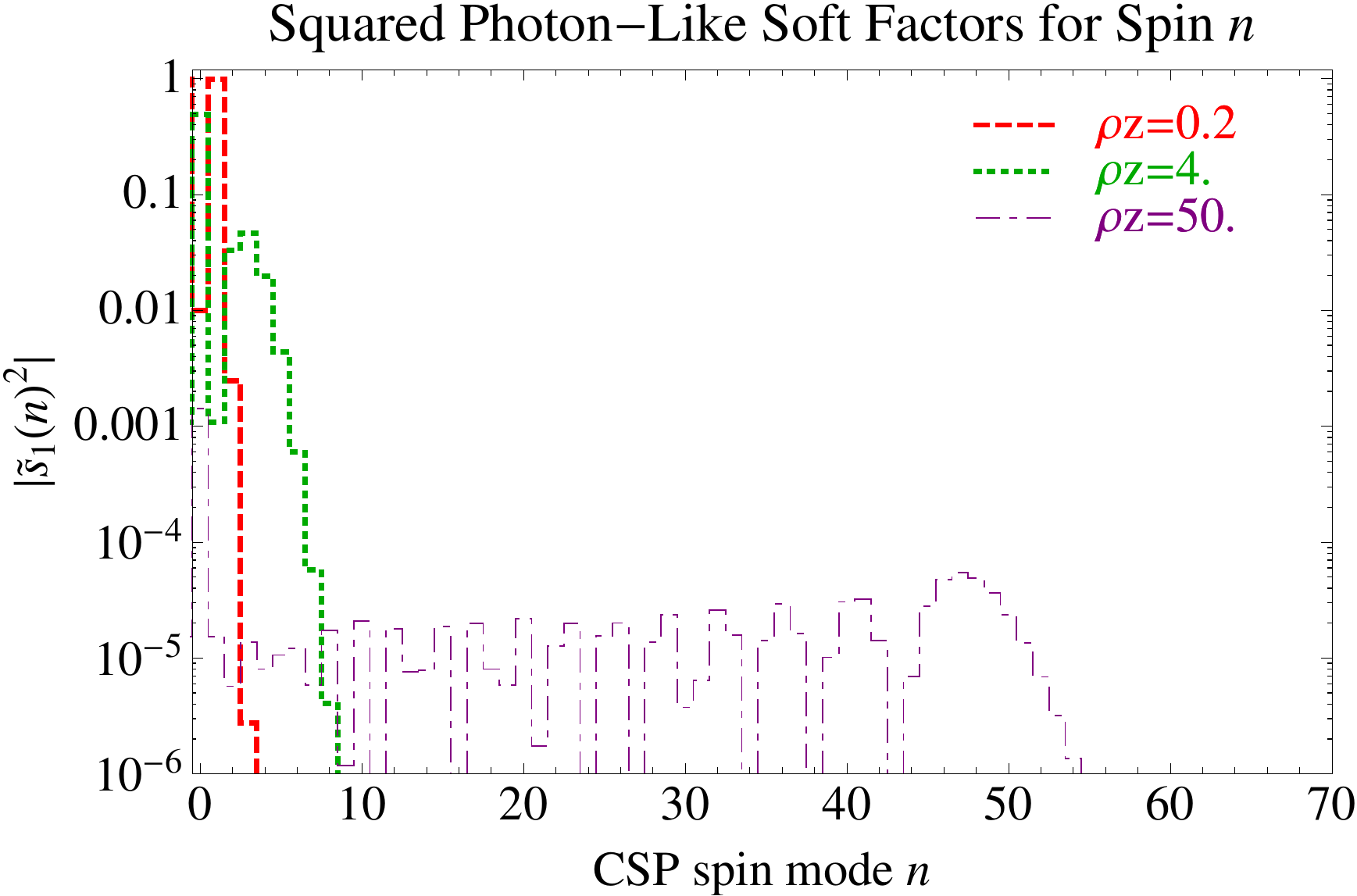}
\caption{\label{fig:photonLikeZplots}
Both plots illustrate on a log scale the scaling of squared photon-like soft factors $|\tilde s_1(\{k,n\},p)|^2$ with $\rho |z| \equiv \sqrt{2} \rho \left|\frac{\epsilon_+(k).p}{k.p}\right|$.  The plots are normalized to the standard QED soft factor for a photon with the same coupling strength.  The left plot focuses on low $n$ modes, while the right plot zooms out to show the large-$n$ scaling of soft factors at large $z$.  For small $\rho |z|$, the amplitudes are sharply peaked at $n=1$ (and -1, not shown).  For large $\rho z$, spin-$n$ soft factors for $n\lesssim \rho |z|$ are $O(\rho |z|)^{-3/2}$ ($O(\rho|z|)^{-1}$ for $n=0$), while $n\gtrsim \rho|z|$ modes are further suppressed.  Negative $n$ modes scale the same way as positive $n$.  The scales are the same as in Figure \ref{fig:scalarLikeZplots}.}
\end{figure}

We can construct a soft factor that is smooth in the $\rho\rightarrow 0$ limit by simply subtracting away the divergent term that cancels (by charge conservation) in amplitudes: 
\be
\tilde s_i^{(1)}(\{k,n\},p_i)  = \frac{- \sqrt{2} e q_i}{\rho} p_i.k \left(\tilde J_n(\rho z_i) - \delta_{n0}\right).\label{eq:CSP1softNCfirst}
\ee
The $\delta_{n0}$ term does not transform like a single-CSP state under Lorentz-transformations, but like a Lorentz scalar, so this soft factor is non-covariant.  It is well-behaved in the $\rho\rightarrow 0$ limit, despite the $1/\rho$ pre-factor, because the factor in parentheses scales as a positive power of $\rho$ for all $n$ ($O(\rho^{|n|})$ for $n\neq 0$ and $O(\rho^2)$ for $n=0$).  The scaling of these soft factors is illustrated in Figure \ref{fig:photonLikeZplots}.
This $\rho\rightarrow 0 $ limit is precisely 
\be
\lim_{\rho\rightarrow 0} \tilde s_i^{(1)}(\{k,n\},p_i) = e q_i \left( \delta_{n,-1} \epsilon_-^*.p_i + \delta_{n,1} \epsilon_+^*.p_i\right).
\ee
To make the correspondence even more manifest, and remove the somewhat misleading $1/\rho$ pre-factor, we can introduce the function  
\be
F_n(w) \equiv - 2 \,\sgn(n) \f{\tilde J_n(w)-\delta_{n0}}{(w)_{\sgn(n)}}, \label{eq:FN}
\ee
where $(w)_{\sgn(n)} = w$ for $n\ge 0$ and $w^*$ for $n< 0$ (in general, we will use $\sgn(0)=+1$ in the following).  In terms of this $F_n$,
\be
\tilde s_i^{(1)}(\{k,n\},p_i)  = e q_i \epsilon^*_{\sgn(n)}.p_i F_n(\rho z_i).\label{eq:CSP1softNC}
\ee
The limiting values $F_{\pm 1}(0)=1$, $F_n(0)=0$ for all other $n$ establish the correspondence.  More precisely, 
the first few $F_n$'s has the small-argument behavior
\bea
F_0(w) & \approx & w/2 (1 - |w|^2/16 + \dots) \\
F_1(w) = F_{-1}^*(w) &\approx& (1  - |w|^2/8 + \dots) \\
F_2(w) = F_{-2}^*(w) &\approx& \;\; -w/4 (1 - |w|^2/12 + \dots).
\eea
The small- and large-argument behavior of $|F_n(w)|$ is easily derived from that of the Bessel functions: 
\be
|F_n(w)| \approx \frac{(w/2)^{||n|-1|}}{|n|!} \qquad \mbox{ for } |w|\ll \max(n,1),
\ee
while at large arguments it is contained in the envelope 
\be
|F_0(w)| < 2/|w| \quad (w=0), \qquad \qquad |F_n(w)| < 2/|w|^{3/2} \quad (w\neq 0). 
\ee
The small-argument approximation is also a bound on $F_n(w)$, which intersects the large-argument envelope at $w\sim n$ for $n\neq 0$ and $w\sim 1$ for $n=0$.

We have seen that the covariant soft factor \eqref{eq:CSP1soft} with generic $q_i$ has correspondence with a momentum-dependent scalar interaction, which violates perturbative unitarity above the cutoff scale $\Lambda \sim \mu$.  However, in the special case that all interactions conserve $q_i$ (i.e. the sum of incoming $q_i$'s equals the sum of outgoing $q_i$'s in any process), the theory has no cutoff and has small-$\rho$ and high-energy correspondence with gauge theory soft factors with gauge coupling $e=\rho/(\sqrt{2}\mu)$.   In this case, the \emph{non-covariant} soft factor \eqref{eq:CSP1softNC} is physically equivalent to \eqref{eq:CSP1soft}, and makes the correspondence more manifest.   The perturbative unitarity argument suggests the intriguing possibility that in CSP theories with a cutoff $\Lambda$, charge conservation (where ``charge'' is defined as the coupling constant of the $n=\pm 1$ modes) may naturally be violated by terms of $O(\rho/\Lambda)$, though the genericity of this effect can only be studied in a full theory.

\subsection{Graviton Correspondence}\label{ss:gravitonCSP}
Analogous arguments apply to the soft factors with pre-factor $f(k.p_i) = g^{(2)}_i (k.p_i)^2$.  The
coefficients $g^{(2)}_i$ have mass-dimension $-3$, and we write them as $g^{(2)}_i = g_i/\mu^3$.  However, it will be convenient to consider a more general $f(k.p_i)$ quadratic in $p_i$: 
\be
s_i^{(2)}(\{k,n\},p_i) = \frac{1}{\mu^3} \left( g_i (p.k)^2 + \frac{g_i'}{4} \rho^2 p^2 \right) \tilde J_n(\rho z_i).\label{eq:CSP2soft}
\ee
Again, the leading interaction at high energies --- the $O(\rho^0)$ term in the emission amplitude for the $n=0$ state --- violates unitarity at a cutoff scale 
$\Lambda \sim \mu$ for generic $O(1)$ $g_i$.  This fastest-growing part of the amplitude is proportional to $1/\mu^3 \sum_i \pm g_i k.p_i^\mu$ ($+$/$-$  sign for outgoing/incoming legs).  The cutoff can only be raised parametrically higher than $\mu$ if $g_i = g$ for all $i$. In this case, the above leading term vanishes by momentum conservation.  Moreover, the $O(\rho^1)$ terms of the $n=\pm 1$ amplitudes, proportional to $g \rho/\mu^3 \sum_i \pm \epsilon_{\pm}(k).p_i^\mu$, also vanish by momentum conservation.  

After these cancellations, the largest amplitudes are for spins 0 and $\pm 2$; these begin at order $1/M_{*} \equiv \rho^2/\mu^3$ and their leading behavior at small $z$ is given by
\wide{
\bea
A(\{k,n=0\},p_1\dots) & = & A_4(p_1\dots)\sum_i \frac{1}{M_*} \frac{- \frac{g}{4} |\epsilon_+^*.p_i|^2 + \frac{g_i'}{4} p_i^2 }{\pm 2 k.p_i} (1+O(|\rho z_i|^2)) \\
A(\{k,n=\pm 2\},p_1\dots) & = & A_4(p_1\dots)\sum_i \frac{1}{M_*} \frac{\frac{g}{8} (\epsilon_\pm^*.p_i)^2}{\pm 2 k.p_i} (1+O(|\rho z_i|^2)).
\eea}
The $n=\pm 2$ amplitude is, at leading order in $\rho z_i$, a gravitational amplitude with Planck scale $M_*$; the first term of the numerator for the $n=0$ mode can be simplified using 
$\epsilon_+^{(\mu} \epsilon_-^{\nu)} = \eta^{\mu\nu} - k^{(\mu} q^{\nu)}$ to the form $- 1/4 (p_i^2 + k.p_i q.p_i)$.  The second term vanishes in the full amplitude by again using momentum conservation.  Thus
\be
A(\{k,n=0\},p_1\dots)  = A_4(p_1\dots)\sum_i \frac{1}{M_*} \frac{1}{\pm 2 k.p_i}\frac{1}{4}(g_i' - g) p_i^2 (1+O(|\rho z_i|^2)).
\ee
For the special choice  $g'_i = g$, the $n=0$ amplitude is governed by the next-leading term, of $O(\rho^2 z^4/M_*)$.  In the  limit $\rho z\rho \ll 1$, this theory approximately reproduces general relativity, with emission amplitudes for the $n$'th mode scaling as $\frac{1}{M_*} (\rho z/2)^{\big||n|-2\big|}/|n|!$.
For generic $g'_i$, the theory also has a scalar with Brans-Dicke-like couplings.  It is not clear, without a full theory, whether the condition $g'_i= g$ arises from minimal theories, or is non-generic in an important way. 

In analogy with \eqref{eq:CSP1softNC}, we may again construct a modified non-covariant soft factor that produces the same amplitudes as \eqref{eq:CSP2soft} and recovers precisely the helicity-2 soft factor in the $z\rightarrow 0$ limit: 
\be
\tilde s_i^{(2)}(\{k,n\},p_i) = \frac{1}{M_* \rho^2} \left((p.k)^2 + \rho^2 p^2/4 \right) \tilde J_n(\rho z_i) - (p.k)^2\left[\delta_{n0} - \delta_{n1} \rho z_i/2 + \delta_{n,-1}\rho z_i^*/2\right]
.\label{eq:CSP2softNC}
\ee
As in the photon-like case, it is straightforward to define ``correspondence functions'' analogous to $F_n(w)$ that smoothly approach $\delta_{|n|2}$ as $w\rightarrow 0$ with increasing powers of $w$ as $n$ gets farther from $\pm 2$.  In terms of these functions, \eqref{eq:CSP2softNC} will look manifestly like a correction to gravitational soft factors. 

Of course, the vanishing in full amplitudes of the non-covariant terms in \eqref{eq:CSP1softNC} and \eqref{eq:CSP2softNC} involves precisely the same cancellation of terms that appeared in Weinberg's soft-emission amplitudes.  Our subtractions are proportional to the  ``pure gauge'' components of a tensor field that are generated by Lorentz transformation.  But the justification for insisting that these terms cancel is entirely different.  In  Weinberg's construction, the constraint was Lorentz invariance; here, covariant soft factors \eqref{eq:CSP1soft} and \eqref{eq:CSP2soft} are easily constructed, but in order to raise or remove a high-energy cutoff, the interaction coefficients must be either conserved (for the $p.k$ pre-factor) or universal (for the $(p.k)^2$ pre-factor), and only in this case do the correspondence limits become photon-like or graviton-like.  

\subsection{General CSP Soft Factors Reconsidered}\label{ss:generalCSP}
One might ask whether this trend continues: can we continue exploiting cancellations to build CSP interactions that couple predominantly to the $n=3$ state in the correspondence limit?  We can certainly construct ``soft factors'' with increasingly high powers  $g_i^{(m)} (p_i.k)^m$ in their pre-factors, but the leading term in an $n=0$ amplitude will always take the form $\sum_i g_i^{(m)} (p_i.k)^{m-1}$.  If the amplitude is to have non-singular support in momentum space (aside from the single momentum-conserving delta-function arising from translation invariance), then for $m>2$ no choice of $g_i^{(m)}$'s will lead to the vanishing of this leading term.  Therefore the ``generic'' CSP soft factor couples dominantly to the spin-0 state; the forms \eqref{eq:CSP1soft} and \eqref{eq:CSP2soft} are very special cases.  

This brings us back to a question about which we have so far been rather glib: whether the ``soft factor'' with generic pre-factor $f(k.p_i)$ is really worthy of the name.  That is, does Lorentz-invariance of the $S$-matrix \emph{require} covariance of the sum over these momentum-dependent soft factors?  A true soft factor is guaranteed to dominate \emph{some} class of amplitudes at small $|{\bf k}|$ --- only in that case is covariance  of the sum of soft factors necessary to ensure covariance of the amplitude. 
Let us consider the growth of the amplitude shown in Figure \ref{fig:single-CSP} in the soft limit for two cases: a scalar-like CSP with arbitrary couplings $a_3$ and $a_4$ and a vector-like CSP interaction with $q_3 = -q_4 = q$.  In the scalar-like case, the emission amplitude is
\be
A_{(0)}(p_1,\dots,p_4,\{k,n\}) = \lambda \left(\frac{a_3}{2 p_3.k} \tilde J_n(\rho z_3) + \frac{a_4}{2 p_4.k} \tilde J_n(\rho z_4)\right) + O((p_i.k)^0),
\ee
while in the vector-like case it is 
\be
A_{(1)}(p_1,\dots,p_4,\{k,n\}) = \lambda \left(-\frac{q}{\sqrt{2} \rho} \tilde J_n(\rho z_3) + \frac{q}{\sqrt{2}\rho} \tilde J_n(\rho z_4)\right)+O((p_i.k)^0). \label{1softExample}
\ee

The behavior of both amplitudes changes crucially between the ``soft correspondence'' regime $\rho |{\bf p}_i| \ll k.p_i \ll p_i.p_j$, where the $\tilde J_n$ is evaluated at small arguments (and is approximately polynomial) and the ``ultra-soft'' regime $k.p_i \ll \rho |{\bf p}_i| \ll p_i.p_j$, where the falling large-argument behavior of $\tilde J_n$ takes over. 
In the ultra-soft regime, $\tilde J_n(\rho z) \sim 1/\sqrt{\rho |z|} \propto \sqrt{k.p_i}$ actually falls as $k$ gets increasingly soft.  The scalar-like amplitude $A_{(0)}$ continues to grow without bound because the propagators grow as $1/k.p_i$ while the Bessel functions fall only as $\sqrt{k.p_i}$.  But the vector-like amplitude $A_{(1)}$ indeed falls in this ultra-soft limit.  ``Internal'' emission contributions to the amplitude, which are \emph{not} enhanced by $1/k.p_i$ poles, could still contribute comparably to the amplitude (and to the total cross-section, which also falls as $\sqrt{k.p_i}$). Therefore, Lorentz-covariance of the external emission components in \eqref{1softExample} is neither necessary to guarantee Lorentz-covariance of the amplitude, nor even sufficient to guarantee Lorentz-covariance of a leading part.  The simplification usually obtained by going to the soft limit has been lost. 

What about the ``soft correspondence'' regime $\rho |{\bf p}_i| \ll k.p_i \ll p_i.p_j < \Lambda^2$, which only exists in a theory with cutoff $\Lambda \gg \rho$?  Here $A_{(1)}(p_1,\dots,p_4,\{k,\pm 1\})$ is well-approximated by helicity 1 soft factors, which \emph{do} grow in the soft limit as $1/k.p_i$.  Indeed, the ``soft'' part of the amplitude \eqref{1softExample} reaches a maximum strength of order $1/\rho$ for $k.p_i \sim \rho |{\bf p}_i|$ ($\rho z_i \sim 1$) before falling off at even softer $k$.  Although we cannot cleanly extract the ``soft part'' of the amplitude as a residue at $k.p_i = 0$, we \emph{can} still expect that it dominates over internal emissions in the soft correspondence regime.  Internal emission contributions to the amplitude are suppressed by propagators $1/p_i.p_j$, and should on dimensional grounds be bounded by $\Lambda/p_i.p_j$.  For momenta near $\Lambda$, this is parametrically smaller than the ``soft'' contributions to the amplitude.  Thus, the scaling of single-emission amplitudes in the ``soft correspondence'' regime provides the justification for demanding covariance of the part of the amplitude obtained from soft factors.  

With the above remarks in mind, we summarize the general soft factor that respects perturbative unitary below a cutoff scale $\Lambda \gg \rho$ and is truly constrained by Lorentz-invariance as 
\be
s_i(\{k,n\},p_i)  = \left[ a_i - \frac{e q_i}{\sqrt{2}\rho} p_i.k + \frac{1}{M_*} \left((p_i.k)^2 - \tfrac{1}{4}\rho^2 p_i^2\right) \right] \tilde J_n(\rho z_i),\label{eq:CSPgeneralSoft}
\ee
where the first term is scalar-like with arbitrary dimension-1 $a_i$, the second term is photon-like and the dimensionless coupling $q_i$ is conserved, and the third term is graviton-like with a universal coupling $1/M_*$ with dimensions of inverse mass.  The latter two terms could equally well be replaced by the non-covariant soft factors \eqref{eq:CSP1softNC} and \eqref{eq:CSP2softNC}.  It is consistent, as far as we can tell from soft factors, for a single CSP with small $\rho$ to mediate helicity 0, 1, and 2-like interactions! These helicities are not singled out by the spectrum (which contains a proliferation of high-spin modes), but by interactions.

A separate but important question is whether some new structure emerges in the ``ultra-soft'' regime.  The correspondence soft factors we have considered, with monomial or binomial $f(p_i.k)$, lead to anarchic ultra-soft spin-basis amplitudes; the ``angle'' basis amplitudes are anarchic for all $k$.  It remains possible that, in some other basis or with another choice of $f(p_i.k)$, the ultra-soft amplitudes display some interesting structure.  This could be a very promising way of untangling the deep infrared physics of CSPs, which is clearly very different from that of either definite-helicity or massive particles.

\section{Infinite-Spin Limit vs. Correspondence Limit}\label{sec:infinite-spin}

The discussion above motivates a physical picture of CSP physics as a generalization of fixed-helicity physics, 
with a physical correspondence in the limit $\rho z \rightarrow 0$.
Another connection often encountered in the literature is of the CSP as a simultaneous high-spin and low-mass limit of a massive particle.  
It's very easy to see that the limit $S\rightarrow \infty$ of a massive particle, with $\rho = M\,S$ held fixed at some finite value, will have
$W^2 = -M^2 S(S+1) \rightarrow -\rho^2$ in the infinite-spin limit (we use $M$ for mass here, to distinguish it from the quantum number $m$).  
The counting of states in the spin-basis is also indicative of a connection between the continuous-spin representations of the \Poincare group 
and high-spin limits of massive particles. Inonu and Wigner formalized this notion in their description of contracting groups and their 
representations \cite{Inonu:1953sp}. Up until now, however, it was unclear whether this relationship was physical or merely group-theoretic.  

Our soft factors allow us to  address this question quantitatively: how are emission amplitudes for a mass-$M$, spin-$S$ limit 
(in the limit of large $S$, and $M$ small compared to particle momenta) related to those for continuous-spin particles?  
For simplicity, we perform the analysis on soft factors rather than full amplitudes, though the conclusions would extend naturally to more 
general amplitudes.  We are reassured to recover some relationship between the massive spin-$S$ soft factors and their CSP counterparts, 
since much of the basic structure of these amplitudes is dictated by Little group symmetry.  
However, it appears to us that the ``infinite-spin'' connection is just group-theoretic -- CSP soft factors have a much better behaved  
analytic structure than the high-spin limit of massive particle soft factors. 
In particular, to obtain a sensible infinite-spin limit, the spin-$S$ soft factors must be multiplied by an $S$-fold pole that, for any finite $S$, grossly 
violates unitarity.  Only when we finally take $S$ to infinity do we recover a CSP soft factor where the $S$-fold pole is replaced by a Bessel function, 
which smoothly cuts off the low-momentum limit of the amplitude.
It would appear unlikely that the physics of an interacting CSP theory is usefully related to the physics of interacting massive particles in 
the infinite-spin limit. 

It is well-known \cite{WeinbergQFT} that massive spin-$S$ fields can be described by traceless and symmetric rank-$S$ 
tensor wavefunctions $A^{{\mu_1}\dots {\mu_S}}$ with $k_{\mu_1}A^{{\mu_1}\dots{\mu_S}} = 0$.  The natural soft factor for 
emission of a spin-$S$ particle of momentum $k^\mu$ off an external momentum $p$ is 
\be
s_S(p,{k,m})= \frac{1}{\Lambda^{S-1}} A_m^{{\mu_1}\dots {\mu_S}} p_{\mu_1}\dots p_{\mu_S},
\ee
where $\Lambda$ is a cutoff scale for the interaction, and $A_m^{{\mu_1}\dots {\mu_S}}$ the wavefunction for the $m$'th spin state.  

We must now specify the basis wavefunctions $A_m$ for the spin-$S$ particle of mass $M$.  To make contact with the massless limit, we will 
want to consider highly relativistic $k$, and work in a basis where $m$ is the eigenvalue of $J_0 = W_0/M$, where $W_0$ is given 
by equation (B6) of \cite{Schuster:2013pxj}.  We note first that 
\be
\epsilon_+, \quad \frac{1}{\sqrt{2}} {J_-}^\mu_\nu \epsilon_+^\nu = -\epsilon_0^\mu, \quad \mbox{ and } 
\quad \frac{1}{2} (J_-^2)^\mu_\nu \epsilon_+^\nu = -\epsilon_-^\mu \label{epsilonBasisSpin1}
\ee
diagonalize $J_0$ and form a canonical basis for the spin-1 $A_m^\mu$ with $m =1$, 0, and $-1$ respectively. 
One can build spin-$S$ wavefunctions out of a symmetric product of $S$ spin-1 wavefunctions, using the definitions
\be
A_S^{{\mu_1}\dots {\mu_S}} = \epsilon_+^{\mu_1} \dots \epsilon_+^{\mu_S}, \quad
A_m^{{\mu_1}\dots {\mu_S}} = \sqrt{\frac{(S+m)!}{(2j)! (S-m)!}} (J_-)^{S-m} A_m^{{\mu_1}\dots {\mu_s}}
\label{buildSpinJ}
\ee
where each copy of $J_-$ is a direct sum of lowering operators $(J_{-,i})^{\mu_i}_{\nu_i}$ acting on the $i$'th index.  
It is now easy to compute $s(p,{k,m})$ from the above, accounting for all combinatoric factors.  In particular, each term in $A_m$ 
can be classified according to the number $k$ of indices acted on by $(J_-)^2$ (for given $k$, it follows that $S-m-2k$ indices 
are acted on by a single $J_-$, and $m+k$ are unaffected), which can range from $\max(0,-m)\leq k \leq \half(S-m)$.  Then 
\bea
s_S(p,{k,m})= \frac{1}{\Lambda^{S-1}} \sqrt{\frac{(S+m)!}{(2j)! (S-m)!}} \sum_k && { \left( \frac{(S-m)!}{2^k} \right) \left( \frac{S!}{k! (m+k)! (S -m -2k)!} \right)} \\
&& \times \left[ (-2\epsilon_-.p)^k (-\sqrt{2} \epsilon_0.p)^{S-m-2k} (\epsilon_+.p)^{m+k} \right],
\eea
where the first factor in parentheses counts the number of ways a given permutation, e.g. the one proportional to 
$\epsilon_-^{\mu_1}\dots \epsilon_-^{\mu_{k}} \epsilon_+^{\mu_{k+1}}\dots\epsilon_+^{\mu_{2k+m}}
\epsilon_0^{\mu_{2k+m+1}}\dots \epsilon_0^{\mu_{S}}$
can arise from action of $J-m$ lowering operators, the second factor in parentheses counting the number of such permutations, 
and the square brackets encoding the resulting contraction with $p_{\mu_1}\dots p_{\mu_S}$.  
This soft factor can be reorganized as
\bea
s_S(p,{k,m})= \frac{1}{\Lambda^{S-1}} && \sqrt{\frac{2^S (S+m)!(S-m)!}{(2S)!}} (-\epsilon_0.p)^S (-e^{i\varphi})^m  \\ 
&& \times \sum_{\max(0,-m)\leq k}^{k \leq (S-m)/2} \frac{(-1)^k}{k! (m+k)!} \left[\frac{S!}{(S-m-2k)!S^{2k+m}}\right] \left(\frac{M S |z|}{2}\right)^{2 k+m},
\eea
where $e^{2i\varphi} \equiv \frac{\epsilon_+.p}{\epsilon_-.p}$ and $|z| \equiv \frac{\sqrt{2} |\epsilon_+.p|}{M \epsilon_0.p}$.

Several simplifications occur in the limit of large $S$ and small mass $M$.  For $m \ll \sqrt{S}$, the overall square-root pre-factor 
becomes an $m$-independent factor $C_S$ (its detailed form is unimportant).  In the small-mass limit, $\epsilon_0.p \rightarrow \frac{1}{M} k.p$ and $|z| \rightarrow \frac{\sqrt{2} |\epsilon_+.p|}{k.p}$.
The sum over $k$ has been written 
in a form closely resembling the Bessel function series expansion.  Indeed, for $2k+m \ll \sqrt{S}$ (corresponding to terms built mainly 
out of $\epsilon_0$), the factor in square brackets approaches 1, so that these terms approach the first few terms of the Taylor 
expansion for $J_m(\rho |z|)$ where $\rho = M S$ (for $m<0$ a change of variables from $k$ to $k+m$ is required to bring these terms into the canonical form).  
The series is dominated for any $|z|$ by terms with $k < |z|/2$, so the whole series can be said to converge to $J_m(|z|)$ at large $S$.  
Then in the large-$S$ and small-$M$ limit,
\be
s_S(p,{k,m}) \sim \frac{C_s p.k^S}{\Lambda^{S-1} M^S}  (-e^{i\varphi})^m 
J_m(\rho |z|).
\ee

This makes it clear that formally,
\be
s_{CSP}(p,{k,m}) = \lim_{S \rightarrow \infty} s_S(p,{k^*,m}) \times \frac{\Lambda^{S-1} M^S}{C_s p.k^S}, \label{eq:inf-spin}
\ee
where $k^*$ is a momentum close to the null $k^\mu$, but with ${k^*}^2 = (\rho/S)^2$.

The fact that the limiting dependence on the spin quantum number $m$ resembles a CSP soft factor is expected, on the basis of group theory.  
Nonetheless, we see that the large-$S$ limit of the massive particle soft factor does not exist in any clear sense, 
because the overall momentum-dependent (but $m$-independent) pre-factor is singular. 
The right-hand side of equation (\ref{eq:inf-spin}) is clearly \emph{not} a quantity that, at any finite $S$, 
could be considered a soft factor -- it has an unphysical order-$S$ pole in $p.k$.  
To us, this suggest that the physical interpretation of a CSP as the infinite-spin limit of a massive 
particle is at best useful for understanding kinematics, but likely not a useful approach for interpreting 
interactions. This conclusion can and will be made sharper in \cite{Schuster:2013pta}.


\section{Helicity Correspondence and Unitarity}\label{sec:unitarity}
We have seen that the soft factors of Table \ref{tab:soft} display correspondence with helicity 0, $\pm 1$, and $\pm 2$ soft factors,
and that these correspondence expressions dominate single-CSP interactions.
Our aim here is to see whether these soft factors can be consistently embedded in a unitary theory, and to explore how correspondence is manifest in this broader context.  

To this end, we consider two other classes of amplitudes: amplitudes with multiple external CSPs in Section \ref{ssec:multiCSP} and amplitudes mediated by CSP exchange in Section \ref{ssec:intermediateCSP}.  At tree-level, unitarity simply amounts to requiring that amplitudes factorize on propagator poles into lower-point sub-amplitudes.  We focus on constructing amplitudes whose factorization limits recover scalar-like CSP soft factors.  In this case, it is rather easy to generalize the soft factors and scattering amplitudes to a consistent set of sewing rules \ref{ssec:sewing}, though the non-uniqueness of the intermediate-CSP sewing rule suggests that a few ingredients are still missing.

There is no reason to expect that vector- or graviton-like counterparts of these amplitudes and rules do not exist, but there are many reasons to expect that they would be more subtle and more cumbersome to obtain by means of unitarity arguments.  Even at tree-level unitarity arguments for scalar theories are simpler than for gauge bosons or gravitons.  Indeed, whereas this factorization follows straightforwardly from scalar Feynman rules, it is quite miraculous even in Feynman-gauge QED: the propagator numerator $g^{\mu\nu}$ is \emph{not} equal to the sum over physical states $\sum_{\lambda = \pm} \epsilon^\mu_\lambda(k) \epsilon^{*\nu}_\lambda(k)$; Feynman rules are only unitarity because the difference between these two expressions, when contracted into the two lower-point factors obtained by Feynman rule construction, vanishes by Ward identities when $k^2=0$.  Although powerful methods exist to compute gauge and gravity amplitudes \emph{via} unitarity, the simplest and most efficient methods like BCFW recursion rely on analytic continuation to complex momenta and on the introduction of a three-point ``amplitude'' that has no real-momentum support.  The complex-momentum technique might be useful for CSPs, but there is no obvious complex-momentum continuation of our soft factor that we have been able to interpret as a three-point amplitude!  Thus, we must confine ourselves to real-momentum arguments -- a more robust starting point, but one that makes gauge theory unitarity arguments fairly involved.  Extending the line of inquiry pursued here to photon- and graviton-like CSP amplitudes or finding obstacles to doing so is of great interest, both theoretically and as a means to build CSP generalizations of known gauge theories and gravity.  We will indicate several subtleties unique to vector- or graviton-like CSPs as they arise in this section.  

\subsection{Correspondence in Multi-CSP Scattering Amplitudes}\label{ssec:multiCSP}

\begin{figure}[!htbp]
\includegraphics[width=0.85\columnwidth]{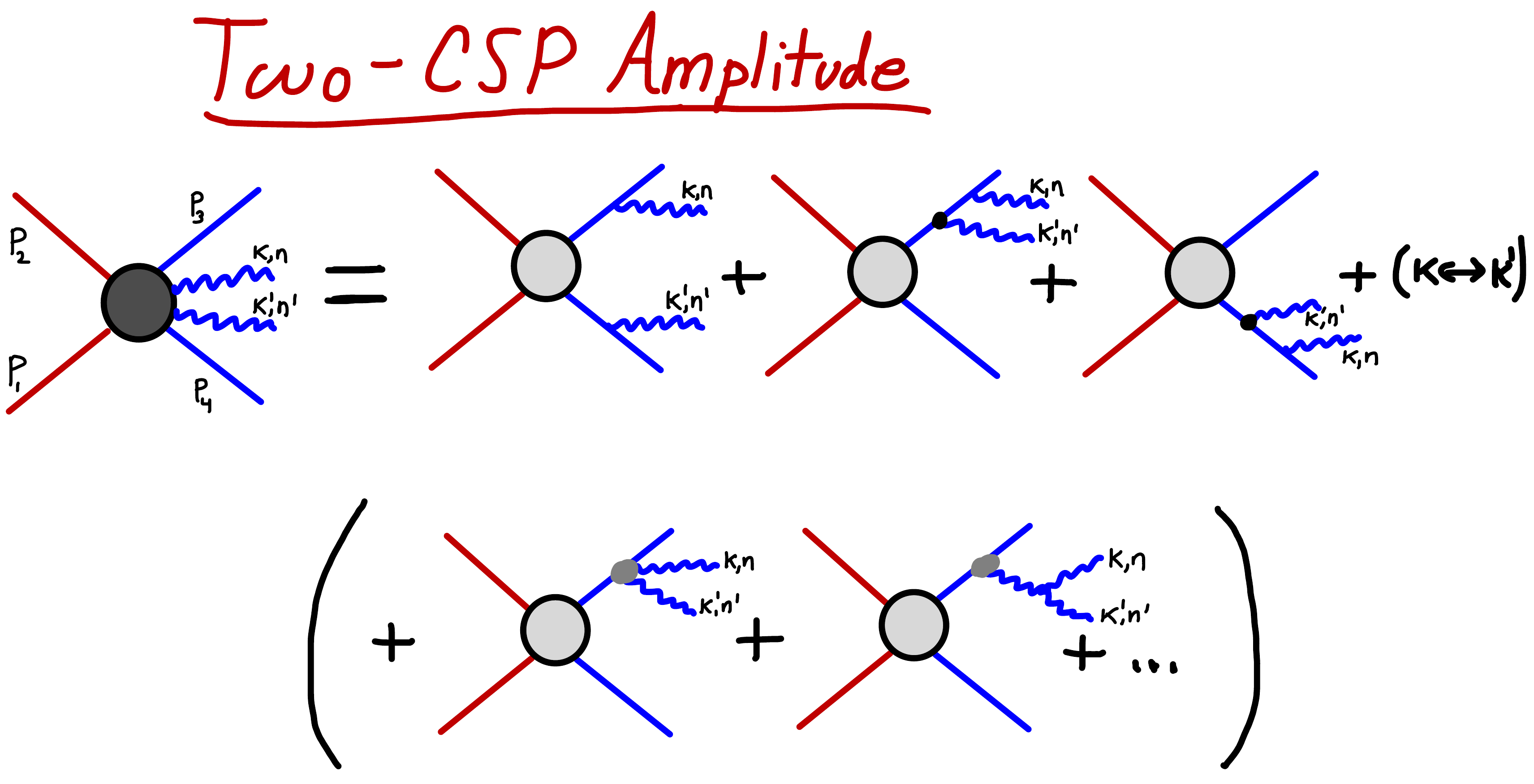}
\caption{The construction of a candidate on-shell two-CSP amplitude is illustrated above. 
CSPs are attached to a parent amplitude $A_4(p_1,p_2; p_3,p_4)$ using the CSP soft factors, 
with appropriate matter propagators included. This example presumes that only the outgoing 
matter legs couple to the CSP. The internal vertex (with black dots) could deviate from the naive soft-factor form, but
we ignore this possibility. The terms in parentheses (with grey dots) describe ``contact''-like structure that could arise.
The final result is the 6-point amplitude $A(p_1,p_2; p_3,p_4,\{k, n \},\{k', n' \})$
used below as an example to investigate certain aspects of CSP interactions.\label{fig:multi-CSP}}
\end{figure}

Constructing a limited class of multi-CSP scattering amplitudes that are Lorentz-invariant and unitary (or at least not conspicuously non-unitary!) is rather 
straightforward for the $m=0$ (scalar-like) soft factors. Consider for example the amplitude for radiation of two CSPs labeled by $\{k,n\}$ and $\{k',n'\}$ off a parent amplitude $A_4(p_1,p_2; p_3,p_4) = \lambda$ (illustrated in Figure \ref{fig:multi-CSP}).
As before, we will start by assuming that only the legs $p_3$ and $p_4$ couple to the CSP.
To maintain unitarity, we expect that when a CSP is radiated off either leg, there must be a propagator factor in the amplitude, which produces a pole when the intermediate line goes off-shell, as well as a soft factor for each emission off an external line.  There is a bit more ambiguity in the internal vertex, when both emissions are off the same line (large dot vertices in Figure \ref{fig:multi-CSP}) --- these vertices could deviate from the soft factor by terms that vanish when either leg is on-shell, but we ignore this possibility as there is no natural deformation of this form.  Neglecting for now the terms in parentheses in Figure \ref{fig:multi-CSP}, we obtain an ansatz six-point amplitude
\bea
A(p_1,p_2; p_3,p_4,\{k, n \}, \{k', n' \}) = \lambda & &
 \frac{a_3 \, s(\{k,n\},p_3)}{((p_3+k)^2+i\epsilon)}\frac{a_4 \, s(\{k', n'\},p_3+k)}{((p_3+k+k')^2+i\epsilon)} \\
& &+\frac{a_4 \, s(\{k, n \},p_4)}{((p_4+k)^2+i\epsilon)}\frac{a_4 \, s(\{k', n'\},p_4+k)}{((p_4+k+k')^2+i\epsilon)}\\ 
& & +\frac{a_3 \, s(\{k, n \},p_3)}{((p_3+k)^2+i\epsilon)}\frac{a_4 \, s(\{k', n\},p_4)}{((p_4+k')^2+i\epsilon)} + (k,n \leftrightarrow k',n'),  \label{eq:multiCSP}
\eea
where $s(\{k,n\},p)$ is the scalar-like soft factor \eqref{eq:CSP0soft}.  It is straightforward to verify that this amplitude continues to yield finite scattering cross-sections, aside from the standard IR divergences. This finiteness is most easily verified in the angle-basis amplitude, obtained either by Fourier transform or by replacing each $s(\{k,n\},p)$ by $s(\{k,\phi\},p)$ from \eqref{eq:CSPsoft} with $f_i = 1$.
Because both $s(\{k,n\},p)$ and $s(\{k',n'\},p)$ have helicity-0 correspondence,  the $n=n'=0$ amplitude is well approximated by a product of scalar amplitudes whenever the $\rho z$ are small (in particular, in the high-energy limits).  Amplitudes for emitting one non-zero $n$ state receive the usual suppression, while amplitudes for $n\neq 0$ and $n'\neq 0$ are suppressed by $O(\rho|z|^{|n|+|n'|})$.  Multi-CSP amplitudes constructed in the same way would follow the same pattern.  

There could also be new terms in the amplitude, involving multiple CSPs at a single vertex, or a CSP self-interaction (the terms in parentheses in Figure \ref{fig:multi-CSP}). Any covariant function without spurious poles would be consistent with the factorization of general multi-emission amplitudes.  Indeed there is considerable flexibility in constructing such a covariant vertex --- for example, the four-point vertex can be obtained from $\psi(\eta,k,\phi)\psi(\eta',k',n')$ by replacing $\eta$ by $k$, the symmetric sum $(p+k+k')+p$ of the two matter momenta at the vertex, or any function thereof.   There is, however, no apparent need to add such terms to our soft factor.  With or without such vertices, there seems to be no obstacle to building a unitary collection of amplitudes for multi-CSP emission.

Each of the ambiguities mentioned above become more important in the case of photon-like (or graviton-like) CSP interactions.  Here, we may approach the problem starting from an expression similar to \eqref{eq:multiCSP}, using the ``subtracted'' (non-covariant but manifestly correspondent) soft factors \eqref{eq:CSP1softNC}).  Just as in QED, the non-covariance in a sum of soft factors vanishes, but the non-covariance in the sum of \emph{products} of soft factors doesn't vanish\footnote{One could also have attempted to build a version of \eqref{eq:multiCSP} out of the un-subtracted photon-like soft factors \eqref{eq:CSP1soft} --- in this case, covariance would be manifest but we would still need to make sure the leading terms cancelled, to preserve unitary factorization of general amplitudes into soft factors.  If there is a field theory for photon-like CSPs, it would likely favor one grouping of Feynman rules or the other, but we don't know which.}. 

In QED, this non-cancellation is cured by the addition of the two-scalar, two-photon vertex; in non-Abelian theories, it further requires a self-interaction vertex.  These vertex rules could have been guessed without reference to an action, simply from unitarity and Lorentz-covariance of amplitudes.  It is clear that CSPs need some similar correction, but the precise form is more difficult to guess.  In contrast to gauge theories, where each vertex is a fixed-rank polynomial in momenta and polarizations (a basis for which can be enumerated), here we face a plethora of candidate vertices. First, the pre-factor of the soft factor $\tilde s_{(1)}(\{k', n'\},p_3+k)$, for example, may be simply $(p_3+k).k'$, but could also receive corrections when both $p_3+k$ and $p_3+k+k'$ are off-shell.  Second, the argument of the four-point vertex Bessel function could depend on several combinations of momenta --- or it might be composed of several terms, each multiplying a different Bessel.  While this freedom makes it easy to find rules consistent with unitarity, Lorentz-invariance, and helicity correspondence for a special case (e.g. CSP pair production), we have not found simple rules that guarantee these properties for \emph{all} multi-emission amplitudes.  It is unclear whether simple rules exist, or whether the theory requires an infinite tower of self-interactions (like gravity or the CSW construction of Yang-Mills theory).  The gauge-theory-like structure for CSPs found in \cite{Schuster:2013pta} suggests that similar machinations might be required here, while ultimately arising from a simple theory.  

\subsection{A Unitary Ansatz for Intermediate-CSP Amplitudes}\label{ssec:intermediateCSP}
In this section, we use unitarity and crossing to construct a family of matter-matter scattering amplitudes mediated by {\it off-shell} CSPs. 
One example, with distinct scalar particles of momentum $a\rightarrow a'$ and $b\rightarrow b'$, is 
\be
{\cal M}_{a+b\rightarrow a'+b'} = 
\f{1}{k^2+i\epsilon} J_0\left(\frac{\rho\sqrt{- (\epsilon^{\mu\nu\rho\sigma} k_\nu p_\rho q_\sigma)^2}}{k.pk.q+p.q k^2}\right) \label{bessel4onshellIntro},
\ee
where $p=a+a'$, $q=b+b'$, and $k=a-a'$. Despite appearances, $J_0(\sqrt{w})$ is analytic on the complex plane ($J_0(w)$ has an even Taylor expansion),
so this amplitude is analytic in momenta, except at isolated points. 
As is now becoming familiar, unitarity arguments do not fully fix the form of the amplitude --- since the amplitude is not a rational function, one can insert terms proportional to $t=k^2$ (the virtuality of the intermediate CSP) in several places without affecting the factorization at null intermediate momentum.  
As a result, the above amplitude is not actually singled out uniquely. Nonetheless, we can see in a straightforward way where this form comes from,
and show that it satisfies the optical theorem at tree-level. 

\begin{figure}[!htbp]
\includegraphics[width=0.85\columnwidth]{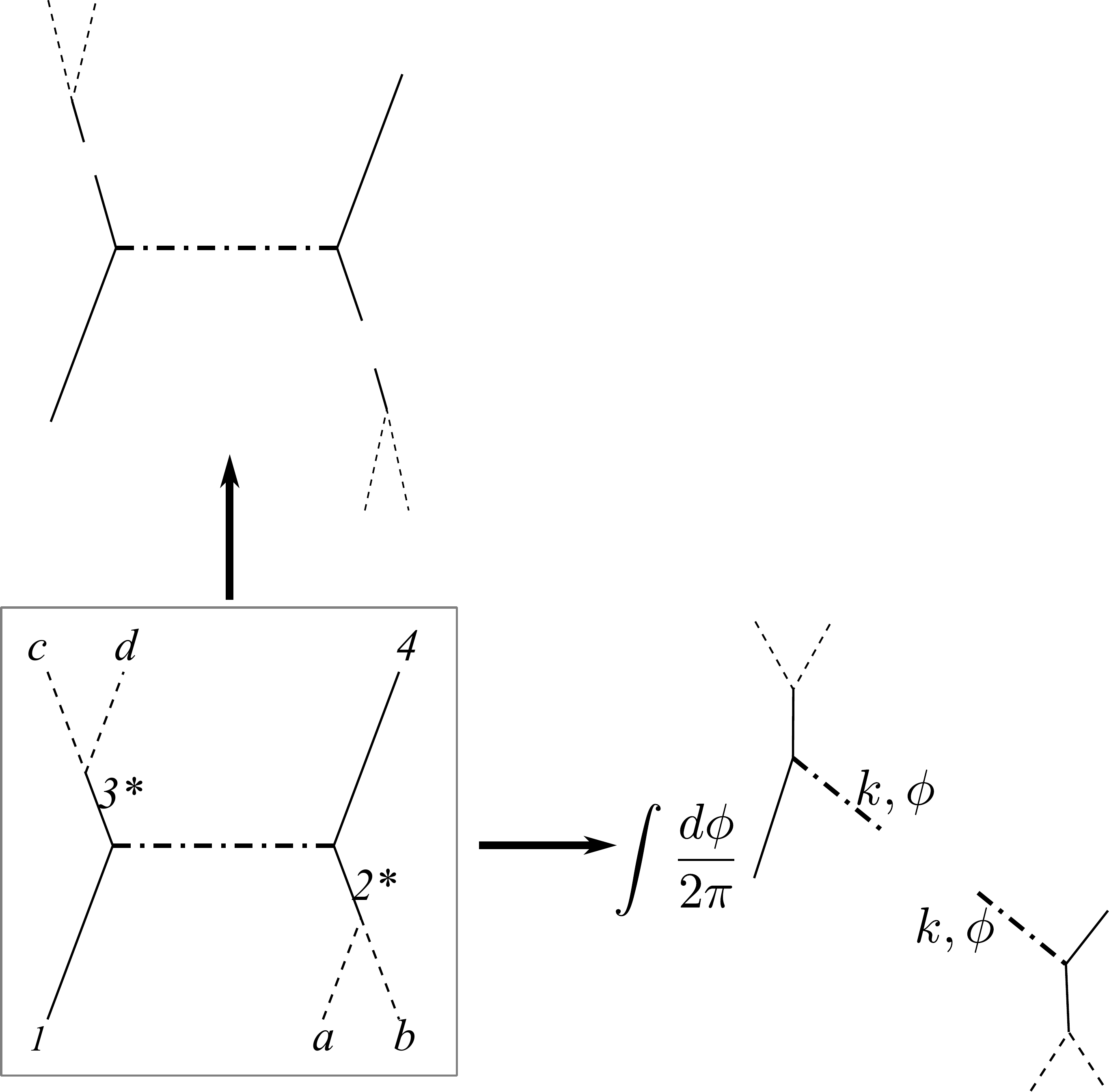}
\caption{The Master Amplitude (bottom-left) used to derive factorization constraints for two-scalar amplitudes mediated by a ``scalar CSP'', and two factorization limits.  The thick dot-dashed line represents an intermediate CSP, solid lines are scalars with a CSP coupling, dashed lines are scalars with no CSP coupling.  In the limit that the pairs $c-d$ and $a-b$ become collinear ($3^*$ and $2^*$ go on-shell) the amplitude factorizes as a product of two splittings and the CSP-mediated amplitude $12 \rightarrow 34$.  In the limit that the CSP goes on-shell, it factorizes into CSP emission and absorption amplitudes. \label{fig:factorizationMasterDiagram}}
\end{figure}

Because unitarity only constrains factorization simply when an intermediate particle goes on-shell \emph{without} becoming collinear (i.e. when its virtuality $k^2$ becomes parametrically smaller than other momentum scales in the problem, including $k.p$), we cannot directly apply unitarity to a massless intermediate state in a $2\rightarrow 2$ amplitude.  Instead, we consider the $3\rightarrow 3$ ``master amplitude'' shown in Figure \ref{fig:factorizationMasterDiagram} (bottom-left), which involves two scalars $\Phi_1$ and $\Phi_4$ that couple to the CSP and four $\phi_{a,b,c,d}$ that do not.  For simplicity we take all particles to be 
distinguishable, so that crossed diagrams need not contribute.  Introducing three-point scalar couplings 
\be
y_3 \Phi_3 \phi_c \phi_d \mbox{ and } y_2 \phi_a \phi_b \Phi_2 \label{three-scalar}
\ee
allows a non-zero amplitude, of the form shown in the diagram.  In the limit $k^2 = (p_c+p_d - p_1)^2 \rightarrow 0$, this should factorize into the product of diagrams in the right of 
Figure \ref{fig:factorizationMasterDiagram}, each of which can be computed by our soft-factor ansatz to obtain an expression (up to terms proportional 
to $k^2$) for the full six-point amplitude.  Taking the limits $(p_c+p_d)^2$ and 
$(p_a+p_b)^2 \rightarrow 0$ of this expression, and re-inserting the possible $k^2$-dependence, we obtain constraints on two-to-two scattering amplitudes.  
In effect, this construction provides a physical regulator of the four-point amplitude by moving the CSP propagator pole away from the collinear region.  

Of course, this construction presumes that the tri-scalar interactions \eqref{three-scalar} can consistently appear in the same theory.  While there is no reason to doubt this assumption for scalar-like CSPs (and it gives a reasonable answer), it would certainly fail for a photon (or a photon- or graviton-like CSP).  In that case, at least one of $\phi_{a,b}$ and one of $\phi_{c,d}$ is charged.  Otherwise, \eqref{three-scalar} would violate charge conservation!  This in turn implies that at least four diagrams would contribute to the six-point master amplitude, with only their sum being gauge-invariant --- a significant complication that we will not tackle here.  We focus here on interactions where in- and out-going particles have equal masses; many of the complications we discuss would actually simplify for unequal masses.  

Let us consider the factorized amplitude on the right side of Figure \ref{fig:factorizationMasterDiagram}.  The four-point amplitudes 
are derivable from the CSP soft factors of Section \ref{ssec:softFactors}:
\bea
{\cal M}_{left} &=&\frac{y_3}{P_3^2 - m^2} c_{1} \exp\left[-i \rho \frac{\epsilon_\phi.p}{k.p} \right] \label{eq:left}\\
{\cal M}_{right} &=& \frac{y_2^*}{P_2^2 - m^2} c_{2}^* \exp\left[+i \rho \frac{\epsilon_\phi.q}{k.q} \right] \label{eq:right}
\eea
where $p=p_1 + P_3$, $q=P_2 + p_4$, and $c_{i}$ denote couplings to the CSPs. We have written the phases in a symmetric 
manner (for $k$ null, $(p+k).k = p.k$ so the above is equivalent to \eqref{eq:CSPsoft}), to assist us in checking crossing later.  

The factorization limit of the six-point amplitude is therefore 
\bea
\lim_{k^2\rightarrow 0} k^2 {\cal M}_{1ab\rightarrow 4cd} & = & S_{3ab} S_{2cd}^* a_{1} a_{2}^* 
\times 
\int \frac{d\phi}{2\pi}  \exp\left[+i \rho \left( \frac{\epsilon_\phi.q}{k.q} - \frac{\epsilon_\phi.p}{k.p} \right)\right] \\
& = & S_{3ab} S_{2cd}^* a_{1} a_{2}^* 
\times 
\int \frac{d\phi}{2\pi}  \exp\left[+i \sqrt{2} \rho \Im\left( e^{-i\phi} \frac{\epsilon_+.A}{k.q k.p}\right)\right],
\eea
where $S_{3ab} = y_3/(P_3^2 -m^2)$, $S_{4cd} = y_4/(P_4^2 -m^2)$, and $A^\mu = k.p q^\mu - k.q p^\mu$.
The $\phi$-dependent phase integrates to a Bessel function
\be
J_0\left( \rho \frac{\sqrt{2}|\epsilon_+.A|}{k.p k.q}\right).
\ee
The quantity $\sqrt{2} |\epsilon_+.A|$ is Lorentz-invariant because $k.A=0$.  To express it in a manifestly covariant form, one can introduce a light-cone vector $Q$ such that $k.Q=1$, $k.\epsilon_{\pm}=0$, in terms of which
\be
g^{\mu\nu} = - \epsilon_+^\mu \epsilon_-^\nu - \epsilon_-^\mu \epsilon_+^\nu +
k^\mu Q^\nu + Q^\mu k^\nu.
\ee
Contracting this identity into $A_\mu A_\nu$, we find
\be
A^2 = -2 \epsilon_+.A \epsilon_-.A = -2 |\epsilon_+.A|^2.
\ee
For the null $k$ that we are considering, we can also write $A^2 = V^2$ with $V^\mu = \epsilon^{\mu\nu\rho\sigma} k_\mu p_\nu q_\sigma$, as these differ only by terms proportional to $k^2$.  Thus, we have 
\be
\lim_{k^2\rightarrow 0} k^2 {\cal M}_{1ab\rightarrow 4cd} =  S_{3ab} S_{2cd}^* a_{1} a_{2}^* 
\times J_0\pfrac{\rho \sqrt{-V^2}}{k.p k.q}.\label{bessel6Point}
\ee
Although writing this in terms of $V^\mu$ seems rather arbitrary, it will be justified later by a rather convenient property of $V^\mu$: as long as $p$, $q$, and $k$ are built from linear combinations of time-like momenta, $V^\mu$ will always be space-like, and $\sqrt{-V^2}$ real.  
Despite appearances, $J_0(\sqrt{w})$ is analytic on the complex plane ($J_0(w)$ has an even Taylor expansion), though it has an essential singularity at infinity.  However, $J_0(\sqrt{w})$ is only bounded for positive $w$ (for negative $w$ it grows exponentially in 
$\sqrt{|w|}$), so it is important that the argument of the square root be bounded from below.

We note that when the intermediate momentum $P_3$ goes on-shell, $p.k=(P_3 +p_1).(P_3 - p_1) = 0$ so the argument of the Bessel function diverges, and the Bessel function itself vanishes! This does not, however, imply that CSP-mediated scattering amplitudes vanish!  We must not forget that \eqref{bessel6Point} is only a $k^2\rightarrow 0$ limit!  

We first consider a simple generalization of \eqref{bessel6Point}:
\be
{\cal M}_{1ab\rightarrow 4cd} = S_{3ab} S_{2cd}^* a_{1} a_{2}^* 
\times 
\f{1}{k^2+i\epsilon} J_0\left(\frac{\rho \sqrt{- V^2 + {\cal F}_4(p,q,k) k^2}}{{k.p\, k.q + {\cal F}_2(p,q,k) k^2}}\right),\label{besselGeneral}
\ee
where ${\cal F}_4$ and ${\cal F}_2$ are Lorentz-invariant polynomials of mass-dimension 4 and 2 respectively (larger classes of deformations are also conceivable, but seem unnecessary).  In this case, the four-point amplitude would have to take the form 
\be
{\cal M}_{12\rightarrow 34} = a_{1} a_{2}^* 
\times 
\f{1}{k^2+i\epsilon} J_0\left(\frac{\rho \sqrt{- V^2 + {\cal F}_4(p,q,k) k^2}}{{k.p\, k.q + {\cal F}_2(p,q,k) k^2}}\right),\label{bessel4}
\ee

We further constrain the functions ${\cal F}$ by the following assumptions:
\begin{enumerate}
\item Invariance under $p_{1,3}\rightarrow - p_{3,1}$ ($s \leftrightarrow u$ crossing), which flips $p\rightarrow -p$ leaving $k$ and $q$ unchanged (and similarly under $q \rightarrow -q$ with $k$ and $p$ unchanged,
\item Invariance under relabeling the diagram to its reflection, i.e. exchanging $p_1 \rightarrow p_2$, $p_3 \rightarrow p_4$, which swaps $p\rightarrow q$, $q \rightarrow p$, and $k\rightarrow -k$, and
\item The amplitude and its crossing variants should all have ${-V^2 + {\cal F}_4(p,q,k) k^2}$ positive-semidefinite.
\end{enumerate}
The first two requirements nearly fix ${\cal F}_2$ --- since $k.p k.q$ is odd under crossing and even under reflection, and $k^2$ is even under both, the unique polynomial ${\cal F}_2$ is $\alpha p.q$.  There is no obvious constraint on the coefficient $\alpha$ from unitarity, except that it should be non-zero so that the argument of the Bessel function does not diverge for generic on-shell momenta.  

As for the numerator, it is easy to verify that ${\cal F}_4 = 0$ satisfies the third requirement, because $V^\mu$ is an epsilon-contraction of physical (null or time-like) momenta, so $V^\mu$ will always be either space-like or null.  One can further check that any polynomial ${\cal F}_4$ satisfying the first two conditions would give rise to $-V^2 + {\cal F}_4 <0 $ either for the $t$-channel amplitude we consider here or one of its crossed variants.  Subject to our assumption \eqref{besselGeneral} for the general amplitude, this fully fixes the numerator of the Bessel argument to be $\rho \sqrt{-V^2}$.

Our simple ansatz amplitude, 
\be
{\cal M}_{12\rightarrow 34} = a_{1} a_{2}^* 
\times 
\f{1}{k^2+i\epsilon} J_0\left(\frac{\rho \sqrt{- (\epsilon^{\mu\nu\rho\sigma} k_\nu p_\rho q_\sigma)^2}}{k.p\, k.q+\alpha p.q k^2}\right),\label{bessel4onshell}
\ee
satisfies all straightforward unitarity checks. Let's go through this explicitly.
The optical theorem 
\be
T_{X \rightarrow Y} - (T_{Y\rightarrow X})^*  = -i \sum_Z T_{X\rightarrow Z} (T_{Y\rightarrow Z})^*, \label{eq:opticalTheorem}
\ee
applied to the matter-matter amplitude above is non-trivial. 
The left hand side of \eqref{eq:opticalTheorem} is proportional to $i\delta(k^2)$ (resulting from the $i\epsilon$ terms),
\be
T_{X \rightarrow Y} - (T_{Y\rightarrow X})^* = -i\delta(k^2)\delta^4(P_X-P_Y)J_0(...). \label{eq:opticalleft}
\ee
The right hand side is obtained by integrating over the phase space (and little group labels) of a product of the soft-factor amplitudes 
$T_{X\rightarrow \{k,\phi \}+...}$ and $(T_{Y\rightarrow \{k,\phi \}+...})^*$ above, 
\bea
-i \sum_Z T_{X\rightarrow Z} (T_{Y\rightarrow Z})^* &=& -i\int d^4k \delta^+(k^2)\delta^4(P_X-P_Z)\delta^4(P_Z-P_Y) \times \nonumber \\
&&\int \frac{d\phi}{2\pi} {\cal M}_{left}(X\rightarrow \{k,\phi \}+...)\times {\cal M}_{right}(\{k,\phi \}+...\rightarrow Y),
\eea 
where the $...$ refers to whatever else is left in the $Z$-state. 
In the case where $X$ and $Y$ each contain three matter legs, $Z$ is a CSP + matter particle state.  We can use \eqref{eq:left} and \eqref{eq:right} for the above ``left'' and ``right'' factors, integrate over phase space and little group labels, and obtain exactly the same as \eqref{eq:opticalleft}, as required by the optical theorem at this order. This is, after all, how the $J_0$ amplitude ansatz was constructed.  

The more subtle case is when $X$ and $Y$ each contain two particles.  The only relevant contribution to the right-hand side at tree-level comes from $Z$ consisting of a CSP state and nothing else.  If either of the two scattered particles has non-zero mass, then the momentum-conserving $\delta$-function has no overlap with $\delta(k^2)$ and both sides of \eqref{eq:opticalTheorem} vanish trivially.  The more subtle case is when all particles are massless. In this case, the product of $\delta$-functions has non-vanishing support on the kinematic configuration where $X$ and $Y$ each consist of multiple massless particles collinear with the CSP momentum $k$.  The argument of the Bessel in \eqref{eq:opticalleft} diverges on the support of $\delta(k^2)$, so that the Bessel function itself vanishes.  Thus, the right-hand side of \eqref{eq:opticalTheorem} must also vanish.  But precisely in this limit, the argument of the Bessel functions in each spin-state's contribution to ${\cal M}_{left}$ and ${\cal M}_{right}$ also diverges.  The natural way to define the soft factors in this limit is by regulating them at slightly off-shell matter momenta.  In this case, the sum over $n$ reduces precisely to the form we found in the six-point amplitude, yielding a single $J_0(\dots)$ that vanishes when the regulator is removed to take the matter legs on-shell.  If one were to do the same calculation in the angle basis, the contribution from each $\phi$ would be a phase of unit norm, but this phase would vary so rapidly near the collinear momentum configuration that we can view it as averaging to zero.  Of course, it would be preferable to frame this argument in a manner where we do not need to invoke a limiting procedure to obtain the right-hand side of \eqref{eq:opticalTheorem} from off-shell objects, but we do not know of a natural way to do so.

In summary, we can obtain 6-point tree-level pure matter amplitudes that factorize correctly into 4-point amplitudes with on-shell CSPs. We expect this to persist for higher-point amplitudes, built for example using the sewing rules described below. But for 4-point pure matter amplitudes, factorization occurs most directly into 3-point {\it soft factors}. If 3-point on-shell CSP amplitudes are defined as the on-shell limit of these soft factors, then they vanish for real momentum. With this interpretation, unitarity in the sense of \eqref{eq:opticalTheorem} is maintained at tree-level. 

\subsection{Scalar-Like CSP Sewing Rules}\label{ssec:sewing}
\begin{figure}[htbp]
\includegraphics[width=0.6\textwidth]{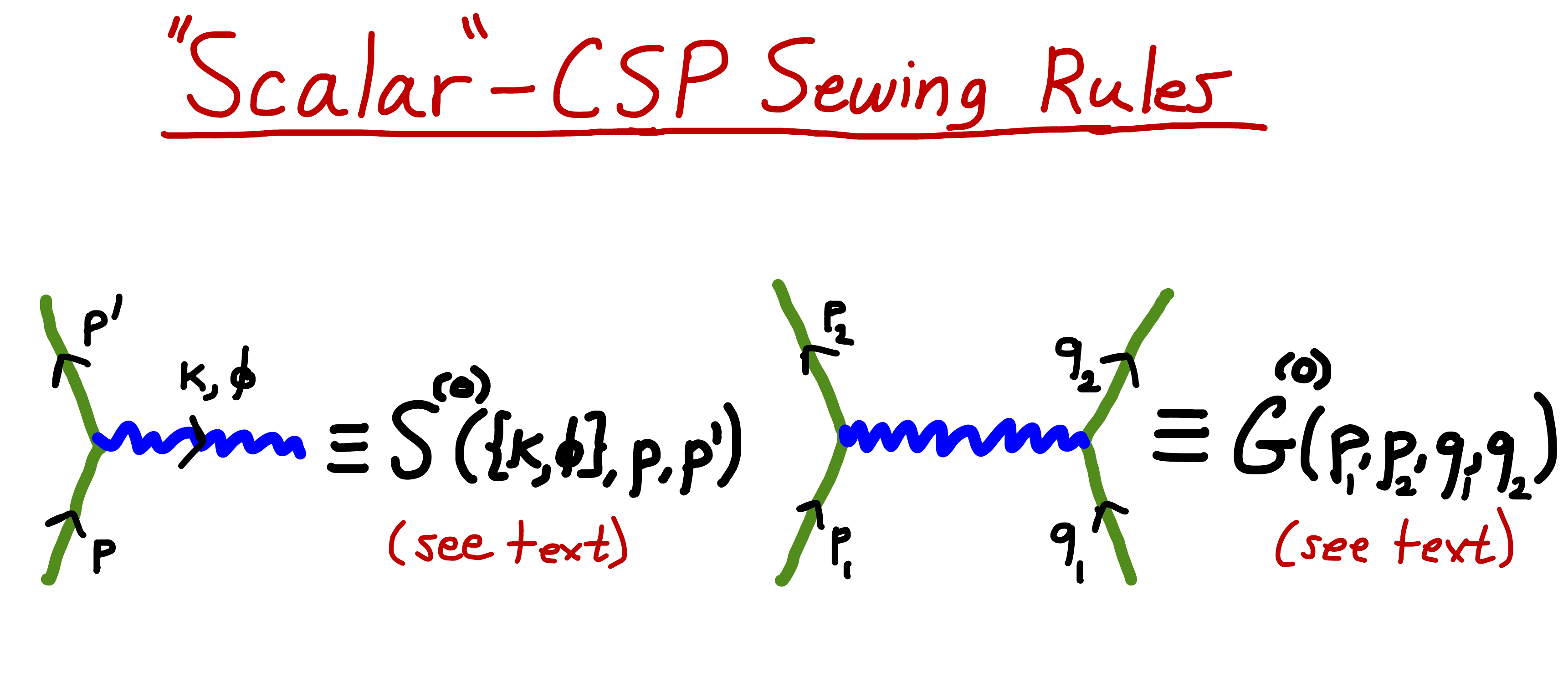}
\caption{A graphical summary of the sewing rules defined in Section \ref{ssec:sewing}.  The left sewing rule \eqref{CSPout} applies to outgoing on-shell CSPs (its conjugate to incoming CSPs) and the right one \eqref{beGen} to intermediate CSPs.  In all cases, the matter four-momenta are arbitrary, and the arrows on matter legs serve merely to define conventions in the formulas.
\label{fig:sewing}}
\end{figure}

The results of this section (with a non-unique modification to \eqref{bessel4onshell} to be described shortly) can be taken as ansatz sewing rules for scalar-like CSPs that appear to be consistent with unitarity at tree-level.  These are summarized in this section and in Figure \ref{fig:sewing}.  

As we showed in \ref{ssec:multiCSP}, the soft factor \eqref{eq:CSP0soft} can be sewn as easily to intermediate scalar matter lines as to external ones.  At tree-level, the resulting amplitudes factorize in a manner consistent with unitarity and preserve the correspondence.  This sewing rule is
\be
s^{(0)}(\{k,n\},p,p')_{out}  = a \tilde J_n(\rho z)\quad s^{(0)}(\{k,\phi\},p,p')  = a \exp\left(i\rho \Im\ [e^{-i\phi} z]\right), \label{CSPout}
\ee
for outgoing CSPs, where
\be
z \equiv\sqrt{2} \epsilon_+^*(k).p/k.p = \sqrt{2} \epsilon_+^*(k).p'/k.p' \quad \tilde J_n (w) \equiv (-1)^n e^{-i n \arg(w)} J_n(|w|).
\ee
For incoming CSPs, the complex conjugate sewing rule 
\be
s^{(0)}(\{k,\phi \},p,p')_{in} \equiv s^{(0)}(\{k,\phi \},p,p')_{out}^*
\ee
should be used (similarly in the spin-basis).  This was shown in \cite{Schuster:2013pxj} (Section IVD) to yield standard crossing relations when analytically continued.  

The generalization of the amplitude \eqref{bessel4onshell} to off-shell external momenta (a correlation function)
is a natural candidate for an intermediate-CSP sewing rule.
However, for off-shell matter momenta there is a region of phase space where all three of $k$, $p$, and $q$ lie in a space-like $(0,3)$ plane.  In this case, $V^\mu$ is time-like, and the argument of the Bessel function in \eqref{bessel4onshell} is unbounded from below.  Indeed, as all momenta become soft in this space-like configuration, \eqref{bessel4onshell} diverges exponentially.
Clearly, something must cut off the resulting exponential growth of the $J_0$ in any physical theory.  There are several candidates, and this difficulty is so far removed from the regime where our unitarity argument is valid that we only mention them briefly.  The sewing rule for attaching a CSP between two off-shell matter legs may be different from the soft factor, though we have found no evidence that this must be the case.  Alternately, \eqref{bessel4} can be generalized in a way that keeps the Bessel function bounded.  One example (keeping terms appropriate for off-shell momenta) is 
\be
G^{(0)}(p_1,p_2,q_1,q_2) = a_{p} a_{q}^* 
\times 
\f{1}{k^2+i\epsilon} J_0\left(\sqrt{\frac{- \rho^2 V^2}{A_4^2 + \beta (\rho^2k^2p^2q^2)^2/A_4^2}}\right), \quad A_4 = k.p\, k.q-\alpha k^2 p.q \label{beGen}
\ee
where $p=(p_1+p_2)/2$, $q=(q_1+q_2)/2$, and any positive $\beta$.  This leaves the form of \eqref{bessel4onshell} unchanged in both high-momentum and soft $k\rightarrow 0$ limits, but ensures that when the argument of the Bessel function becomes negative (which implies $|V^2| < |k^2 p^2 q^2|$), the argument of the Bessel cannot become large.  At the same time, the arbitrariness of \eqref{beGen}, with its complicated form and two free coefficients, makes it clear that a better physical interpretation for these objects is badly needed. 
It is provided merely as a demonstration that a bounded sewing rule can be consistent with tree-level unitarity, and can be used for 
constructing n-point CSP amplitudes for a simple theory with a single CSP interacting with a scalar. 
General amplitudes for a given process are constructed by summing over all allowed channels (graphs)
using the rules illustrated in Figure \ref{fig:sewing} to compute each graph. 

It would be very interesting to examine quantum properties of a theory built from these sewing rules, both to see whether it is unitary and physically reasonable at a quantum level and to see whether any physical arguments can further constrain the form of \eqref{beGen} -- this is an important open problem.  
It would also be very interesting to see whether analogous (but presumably more complicated) sewing rules exist for photon- and graviton-like CSPs.  

\section{CSP Interactions with Gravity} \label{sec:CSPgravity}
Another classic and powerful constraint on high-helicity theories is based on the difficulty of coupling high helicity particles to gravity.  One may worry that similar obstructions arise in coupling CSPs to gravity.  We have endeavoured to exclude CSP couplings to gravity by several standard arguments, and have so far been unsuccessful.  All the same, we lack a model of fully consistent gravitational interactions of CSPs.  In this section, we show how some of the widely known no-go arguments against high helicity don't quite work for CSPs, and suggest some more subtle but potentially severe constraints on CSP interactions with gravity.  We close with a roadmap of distinct possibilities for CSP gravitation.  We hope these remarks will inspire more conclusive work to constrain or develop each of these options.  

At the outset, we distinguish two very different classes of potential obstructions to gravitational CSP couplings: one formal, and one physical.  The first question concerns the existence of Lorentz-covariant graviton-CSP interactions.  Though the simplest generalization of the Weinberg-Witten theorem \cite{Weinberg:1980kq} fails to exclude such interactions, further obstructions may exist.  The second concerns problematic physical effects of gravitational CSP interactions.  The simplest such problem would be if cross-sections $\sigma(e^+e^-\rightarrow CSP\, CSP)$ through an off-shell graviton formally diverge, as intuition motivated by the equivalence principle in the spin-basis would suggest.  We will consider each potential problem in turn. 

The Weinberg-Witten theorem \cite{Weinberg:1980kq} is a remarkably simple argument forbidding a non-vanishing and covariant energy momentum tensor $T^{\mu\nu}$ for massless particles of helicity $>1$.  
Essentially, the argument considers the action of rotations on matrix elements $M^{\mu\nu} \equiv \bra{p',h}T^{\mu\nu}\ket{p,h}$ in a brick-wall frame ${\vec p} = -{\vec p'}$.  Lorentz-invariance dictates that rotations about the ${\vec p}$ axis must leave $M^{\mu\nu}$ invariant, as they do not change $p$, $p'$, or $h$.  On the other hand, these rotations re-phase both single-particle states by $e^{ih\theta}$ (with the same sign) and components of $T^{\mu\nu}$ by $e^{im\theta}$ with $|m|\leq 2$.  For $h>1$, the net transformation $e^{i(2 h + m) \theta} \neq 1$ is inconsistent with the invariance of $M^{\mu\nu}$, unless $M^{\mu\nu} = 0$.

Unlike the high-helicity case, one can easily construct a non-zero, Lorentz-covariant, and conserved tensor matrix element for CSPs. For example: 
\bea
\bra{p',\phi'}T^{\mu\nu}(k)\ket{p,\phi} & = & (p^\mu p'^\nu+p'^\mu p^\nu- p.p' g^{\mu\nu}) e^{i \rho \left( \f{\epsilon_{\phi'}(p').k}{p'.k}- \f{\epsilon_{\phi}(p).k}{p.k} \right)},\\
& = & (p^\mu p'^\nu+p'^\mu p^\nu- p.p' g^{\mu\nu}) e^{-i \rho \left( \f{\epsilon_{\phi'}(p').p  + \epsilon_{\phi}(p).p'}{p.p'} \right)},
\eea
or in the spin basis
\be
\bra{p',n'}T^{\mu\nu}(k)\ket{p,n} = (p^\mu p'^\nu+p'^\mu p^\nu- p.p' g^{\mu\nu}) \tilde J_{n'}\p{\rho \f{\epsilon_+(p').p}{p.p'}} \tilde J_n^*\p{\rho\f{\epsilon_+(p).p'}{p.p'}}.\label{candidateMmunu}
\ee
These matrix elements furnish appropriately leading-order covariant gravity scattering amplitudes; their contraction into the graviton momentum $k^\mu = p'^\mu - p^\mu$ vanishes, ensuring gauge-invariance of this leading-order amplitude.
While this is encouraging, it certainly does not guarantee that CSPs can couple consistently to gravity at all orders -- 
it is not clear that the above can be interpreted as matrix elements of an energy-momentum tensor that is conserved as an operator.

Though the relevance of \eqref{candidateMmunu} to gravity is unclear, we mention for completeness several notable physical properties.  For $k^2 \gg \rho^2$, the matrix element \eqref{candidateMmunu} approaches $\delta_{n0} \delta_{n'0} T_{\mu\nu}^{scalar}$, so that it has helicity-0 correspondence, and there may be similar operators with helicity-1 correspondence.  At the same time, it has a new and somewhat worrisome property: exchange of a single soft graviton with $k^2 \ll \rho^2$ would maximally mix different spin states.  This unphysical effect could potentially be ameliorated by a very small $\rho$, and we do not know whether it persists in a full eikonal calculation of CSP scattering in a weak gravitational field.  
In any case, total scattering cross-sections obtained from \eqref{candidateMmunu} are finite, as are the production cross-sections that would be obtained from the crossed matrix element. 

This brings us to the second question about \eqref{candidateMmunu}, and about CSP coupling to gravity more generally.  Does an amplitude like \eqref{candidateMmunu} satisfy the equivalence principle?  And if not, would equivalence-principle-respecting amplitudes necessarily lead to divergent rates for gravitational processes, from perturbative CSP pair production to Hawking emission?  

A sharp way of framing the equivalence principle for CSPs in terms of an $S$ matrix is to seek Lorentz-covariant amplitudes with at least one graviton and one CSP among their external states and study their properties (the simplest such amplitude would be for annihilation of two scalars into a graviton and a CSP).  The equivalence principle in this context usually means that each state should have unit coupling to the graviton \cite{Weinberg:1964ew}. It is not clear how this requirement should be modified when dealing with CSP matter, which introduces a CSP propagator.  Though this argument is not a direct test of \eqref{candidateMmunu} (the amplitude in question involves an off-shell CSP and on-shell graviton, while \eqref{candidateMmunu} is only defined for on-shell CSPs), it may shed some light on potential obstructions to the equivalence principle for CSPs.  

The form of \eqref{candidateMmunu} also underscores why the worry about divergent cross-sections is premature.  While we might expect $s$-channel graviton matrix elements $M^{\mu\nu} = \langle T^{\mu\nu}\ket{\{k,n\},\{k',n'\}} \sim k^\mu k'^{\nu} \delta_{nn'}$ based on the equivalence principle, leading to a divergent sum over $n$, this simple guess is not Lorentz-covariant --- modifying it by adding $J_n$ form factors as in  \eqref{candidateMmunu} would lead to finite cross-sections, as already noted.  And such form factors need not be at odds with the equivalence principle.  After all, the gravitational Hydrogen-Antihydrogen production cross-section is finite, despite the infinite multiplicity of hydrogen bound states below $m_e+m_p$.  That puzzle is resolved by form factors with a simple physical interpretation: gravitons have exponentially suppressed couplings to the tower of increasingly de-localized bound states.  Though we do not have the same sharp physical picture in the case of CSPs, the very specialized non-local character of the $J_n(\rho z)$'s on which CSP form factors depend does seem consistent with a similar effect.   While this is still poorly understood, it is not a uniquely gravitational problem.  

All of these questions can be framed far more precisely in a field-theory context \cite{Schuster:2013pta}.  In this case, we can identify several possible resolutions.  The first and most concrete is based on coupling of a helicity-2 $h^{\mu\nu}$ to the canonical Belinfante energy-momentum tensor $\Theta^{\mu\nu}$ in a CSP gauge field theory.  If this $\Theta^{\mu\nu}$ is suitably invariant under the CSP gauge symmetry, then it forms the basis for a formally consistent coupling to gravity, and allows a sharp evaluation of physical questions like production cross-sections.  If no suitably invariant $\Theta^{\mu\nu}$ exists, it rules out the standard geometric interpretation for CSP couplings to gravity.  It might still be possible that helicity-2 gravity could couple to a conserved and covariant tensor $T^{\mu\nu}$ that is \emph{not} the Belinfante stress-energy tensor.  In the $\rho=0$ limit, it is certainly consistent for only the correspondence states to interact gravitationally --- the accompanying tower of non-gravitating high helicity states would pose no problems, since they have no other interactions.  A $\rho\neq 0$ generalization of this physics would be surprising, but we do not know of general arguments against this possibility in systems with infinitely many species.  
A final possibility is that CSPs may more naturally couple to gravity mediated by a CSP with helicity-2 correspondence, rather than by exact helicity-2 gravitons.  This would at least be an economical use of CSPs.  If such interactions can be realized (even in the simplest form of a single self-interacting ``CSP graviton''), their non-linear structure would likely generalize in an interesting way the curved space interpretation of general relativity!  This possibility heightens the motivation to better understand multi-CSP interactions.  

\section{Conclusions and Future Directions}\label{sec:conclusion}

In \cite{Schuster:2013pxj}, we presented new wave equations and wavefunctions for continuous-spin particles, and identified the unique family of Lorentz-covariant and smooth soft factors for the interaction of continuous-spin particles with matter.  In this paper, we have demonstrated that the soft factors exhibit a \emph{helicity correspondence} at high energies, and that the simplest ``scalar-like'' soft factors can be sewn together into consistent tree-level amplitudes for multi-CSP emission and for CSP-mediated scattering, and are consistent with gravity at leading order.

We first showed that if a CSP theory has a high cutoff $\Lambda \gg \rho$, three specific forms of soft factor yield cutoff un-suppressed (or in the last case, weakly suppressed) interactions.  These three soft factors exhibit a \emph{helicity correspondence}: amplitudes for emission of a CSP whose energy $E$ exceeds $\rho v$ are dominated by a single spin mode, this spin mode's amplitudes are approximated by helicity amplitudes up to corrections of order $(\rho v/E)^2$, and production of other spin modes is suppressed by increasing powers of $\rho v/E$.  Here $v$ denotes the velocity of the emitter in the frame where the Little Group rotation coincides with a Lorentz rotation, i.e. where $\epsilon_\pm^{0}=0$.

We demonstrated that multi-CSP-emission amplitudes that continue to exhibit scalar-like correspondence can be obtained by sewing together CSP soft factors.  In a similar vein, we use unitary factorization of CSP-mediated matter scattering to motivate a new sewing rule for off-shell CSPs.  Unitarity does not fully constrain the matter-matter scattering amplitude or sewing rule, and appears to be consistent with several deformations, all of which maintain scalar-like correspondence.  
We also exhibited a consistent leading-order CSP coupling to gravity that also exhibits scalar-like correspondence.  Here, however, the correspondence is controlled by the soft graviton momentum rather than the CSP momentum.  It is an open question whether and how these results generalize to photon- and graviton-like correspondence.

Taken together, these results motivate speculation that theories of interacting CSPs might exhibit all three types of correspondence, either simultaneously or on different ``branches'' of the theory; one theory might be approximately described by scalar interactions at high energies, another by gauge bosons, and a third by gravity.   This is of course an exciting possibility --- it suggests that CSPs might furnish natural generalizations of QED and gravity, in which the flat-space photon (graviton) remains massless but does not have a Lorentz-invariant helicity.  Some constraints on this possibility and avenues for testing it will be discussed in \cite{SchusterToro:ph,Evans}. 

It is likely that many of the most basic questions we would like to answer about CSPs --- including some with obvious phenomenological implications, like the modification of force-laws at distances of $O(\rho v)^{-1}$, the possible existence of a massive phase, and even CSP ``electrostatics'' ---  are more efficiently described in a field theory that elucidates the classical limit of these theories.  
Nonetheless, many basic theoretical questions that can and should be approached from an S-matrix viewpoint are left unanswered.  Several of these are highlighted below, with an emphasis on those that can help to untangle the consistency and behavior of photon-like and graviton-like CSP theories.

\begin{itemize}
\item {\bf Scattering Mediated by Off-Shell CSPs:} A variety of CSP-mediated scattering reactions warrant further investigation.  In \S\ref{ssec:intermediateCSP} we constructed a six-point amplitude that factorizes into scalar-like soft factors at real momenta, then used it to infer two-to-two scattering amplitudes consistent with real-momentum unitarity.  It would be very valuable to construct analogous amplitudes that factorize to photon- or graviton-like soft factors, or to prove that no such amplitude exists.   Besides enabling the study of CSP force laws, this would also provide clues for how to naturally describe CSP degrees of freedom off the mass shell.  The technique of \S\ref{ssec:intermediateCSP} can be generalized to these cases, but charge conservation implies that multiple channels must contribute to the six-point amplitude.  This introduces subtleties to the  photon- and graviton-like arguments that were absent from the scalar-like case.  It would be very useful to develop further tools for constraining these amplitudes, which do not require the ``physical regulator'' of the six-point amplitudes or simplify computation of this six-point amplitude.  One possible strategy is using matter-leg shifted BCFW \cite{Britto:2004ap,Britto:2005fq} or Risager \cite{Risager:2005vk} deformations to expose poles at complex momentum.  This requires a prescription for handling the complexified soft factor, and a generalization of BCFW recursion relations to amplitudes with isolated essential singularities.  
\item {\bf Multi-CSP Amplitudes:} The single-emission soft factors for scalar-like CSPs could be readily interpreted as sewing rules to build multi-CSP emission amplitudes (\S\ref{ssec:sewing}) with a product structure much like that of ordinary scalar amplitudes, which automatically factorize in a manner consistent with both unitarity and helicity correspondence.  The analogous task for photon-like amplitudes is harder.  Here, the soft factors cannot be interpreted as a \emph{complete} set of sewing rules, but appear to require correlated contact interactions to simultaneously preserve the Lorentz-covariance of multi-CSP amplitudes or their high-energy boundedness.  This is not surprising, as scalar QED (to which the CSP theory would have to reduce) has four-point vertices required to preserve Lorentz-invariance and unitarity.  It would be interesting and very useful to seek a systematic set of rules for higher-point photon-like CSP amplitudes (and likewise for the graviton-like amplitudes).  The structure of these rules may also shed light on the extent to which CSP theories are less local than ordinary gauge theories.  
\item {\bf Self-Interacting CSPs:} Motivated by the possibility of a general helicity correspondence, it would be interesting to construct 
self-interaction amplitudes for CSPs, and in particular to seek amplitudes that approach non-abelian gauge theory amplitudes in the $\rho\rightarrow 0$ correspondence limit. Similar constructions for graviton-like self-interaction amplitudes would also be very interesting, as would interactions of a graviton-like CSP with scalar- or photon-like CSPs.  It would be particularly interesting to see whether these ameliorate the ``democratization'' of spin-basis states by soft scattering off a graviton through the matrix element \eqref{candidateMmunu}.
\item {\bf Fermionic and Supersymmetric Generalizations:} This paper and \cite{Schuster:2013pxj} have focused principally on CSP-scalar amplitudes where the CSP has integer spins.   It would be valuable to generalize the single and multi-CSP amplitudes constructed in this paper to 
include coupling to fermionic matter.  Fermionic matter may also have new types of interactions with ``double-valued'' continuous-spin representations along the lines considered in \cite{Iverson:1971vy}.  Continuous-spin representations also have supersymmetric generalizations, discussed in \cite{Brink:2002zx}, which may permit non-trivial interactions that generalize the ones found in this paper.   The additional symmetry constraints on supersymmetric CSP amplitudes might facilitate progress in understanding CSP interactions more generally.
\item{\bf Low-Energy Behavior:} The correspondence highlights a remarkable simplification of CSP interactions at energies $E \gg \rho$.  In the spin-basis, CSP amplitudes in this high-energy limit approach helicity amplitudes.  It is not clear what structure, if any, CSP soft factors have in the ultra-low-energy limit $E\ll \rho$.   Spin- and angle-basis matrix elements are both fairly anarchic in this regime, making the physical picture of processes with characteristic energies $E<\rho$ very unclear. The uniform large-argument behavior of $J_n$ with different $n$ suggests that perhaps there is some other basis in which the dynamics of this regime is hierarchical.  Finding or excluding such a structure would shed light on the puzzling long-distance physics of CSPs. 
\end{itemize}

\section*{Acknowledgments}
We thank 
Haipeng An,
Nima Arkani-Hamed,
Cliff Burgess, 
Freddy Cachazo,
Yasunori Nomura,
Maxim Pospelov, 
Yanwen Shang,
Carlos Tamarit,
Mark Wise,
and Itay Yavin
for helpful discussions on various aspects of this physics.
We additionally thank Jared Kaplan for useful feedback on the manuscript. 
Research at the Perimeter Institute is supported in part by the Government of Canada through NSERC and by the Province of Ontario through MEDT.  

\appendix
\bibliography{CSP_correspondence}
\end{document}